\documentclass[fleqn,usenatbib]{mnras}

\usepackage{newtxtext,newtxmath}

\usepackage[T1]{fontenc}

\DeclareRobustCommand{\VAN}[3]{#2}
\let\VANthebibliography\thebibliography
\def\thebibliography{\DeclareRobustCommand{\VAN}[3]{##3}\VANthebibliography}

\usepackage{graphicx}	
\usepackage{amsmath}	
\usepackage{fix-cm}
\usepackage{booktabs}

\title[Spectral Variability in NGC 4490 ULX-8]{Probing spectral variability in NGC~4490 ULX-8 over 24 years of \textit{XMM-Newton}, \textit{Chandra} and \textit{Swift}-XRT observations}

\author[T. Vashisht et al.]{
Tarang Vashisht$^{1}$\thanks{E-mail: tarang0112358@gmail.com},
Aru Beri$^{1,2,3}$,
Tanuman Ghosh$^{4}$,
Aman Upadhyay$^{5}$,
Vikram Rana$^{5}$
\\
$^{1}$Department of Physical Sciences, Indian Institute of Science Education and Research (IISER) Mohali, Punjab 140306, India\\
$^{2}$School of Physics \& Astronomy, University of Southampton, Southampton, Hampshire SO17 1BJ, UK\\
$^{3}$Indian Institute of Astrophysics, Koramangala II Block, Bangalore-560034, India\\
$^{4}$Inter-University Centre for Astronomy and Astrophysics, Ganeshkhind, Pune 411007, India\\
$^{5}$Astronomy and Astrophysics, Raman Research Institute, Sadashivanagar, Bangalore 560080, India\\
}

\date{Accepted XXX. Received YYY; in original form ZZZ}

\pubyear{\the\year{}}

\begin{document}
\label{firstpage}
\pagerange{\pageref{firstpage}--\pageref{lastpage}}
\maketitle

\begin{abstract}
We present a spectral variability study of the ultraluminous X-ray source NGC~4490~ULX-8 based on 14 \textit{Chandra}, 6 \textit{XMM-Newton} and 19 \textit{Swift}-XRT observations obtained between 2000 and 2024. The X-ray spectra are modelled using absorbed power-law and absorbed multicolour disc blackbody models. The best-fit photon indices span $\Gamma \simeq 0.9$--2.7, while the inferred inner disc temperatures lie in the range $kT_{\mathrm{in}} \simeq 1.0$--1.6~keV.
We detect pronounced long-term variability in the unabsorbed X-ray luminosity on multi-year timescales, while variability within individual observations is comparatively modest. A Hardness–Intensity Diagram of the source shows no clear transition between hard and soft states; however, two recent observations taken on 2022 December 1 and 2024 May 4 show a sharp increase in brightness. The spectra across all observations are dominated by smooth, single-component curvature in the 0.3--10~keV band, consistent with the broadened-disc regime of ultraluminous X-ray sources.
A correlation analysis reveals a weak positive $L_{\mathrm{X}}$–$\Gamma$ trend that remains statistically supported after controlling for absorption-related degeneracies, indicating that it is not driven solely by fitting covariance. The $L_{\mathrm{X}}$--$T_{\mathrm{in}}$ relation is only weakly constrained, but remains compatible, within uncertainties, with both thin-disc and slim-disc scalings.
Using disc parameters derived from higher-quality \textit{XMM-Newton} spectra, we obtain model-dependent estimates of the characteristic inner disc radius and compact-object mass as functions of inclination and spin. The reported results are consistent with a stellar-mass black hole accretor operating at or near the Eddington limit.
\end{abstract}




\section{Introduction}
\label{Introduction}
Ultraluminous X-ray sources (ULXs) are point-like, non-nuclear extragalactic compact objects with isotropic luminosities ($L_{\mathrm{X}}$) exceeding $10^{39}\ \mathrm{erg\ s^{-1}}$ \citep{Kaaret2017, King2023}, roughly the Eddington limit for a 10\,M$_{\sun}$ black hole (BH). ULXs consist of a compact object, either a BH or a neutron star (NS) accreting from a companion star.
They were initially hypothesized to host intermediate-mass black holes (IMBHs; $10^{2}$–$10^{5}$ M$_{\sun}$) accreting at sub-Eddington rates \citep{Colbert1999, Miller2004}, an interpretation invoked to explain ULXs in general and, in particular, the more extreme hyperluminous X-ray sources (HLXs) with luminosities exceeding $10^{41}\ \mathrm{erg\ s^{-1}}$ (e.g. \citealt{Webb2010}, \citealt{MacKenzie2023}). While HLXs are still widely regarded as strong IMBH candidates, growing evidence favours the interpretation that most ULXs are stellar-mass BHs or NSs accreting at near- to super-Eddington rates, facilitated by optically thick outflows or geometrical beaming \citep{Fabrika2015}. This has been supported by dynamical mass estimates, detection of ULXs hosting stellar-mass BHs \citep{Shen2015, Motch2014, Fabrika2015} and the discovery of more than 10 pulsating ULXs (pULXs), definitively indicating NS accretors, such as M82 X-2 \citep{Bachetti2014}, NGC 7793 P13 \citep{Fuerst2016, Israel2017}, NGC 5907 ULX1 \citep{Israel2017a}, NGC 300 ULX1 \citep{Carpano2018} and NGC 1313 X-2 \citep{Sathyaprakash2019}.

NGC 4490 is a late-type spiral galaxy currently interacting with the irregular galaxy NGC 4485. A distance of 7.8 Mpc, based on the Tully–Fisher (TF) relation \citep{Tully1988a}, has been widely adopted in the literature for the system. More recent tip of the red giant branch (TRGB) measurements using \textit{Hubble Space Telescope} (\textit{HST}) data place NGC 4490 at 6.5 Mpc and NGC 4485 at 8.8 Mpc \citep{Sabbi2018a}. These distances imply a three-dimensional separation of $\sim$ 2.3 Mpc between the galaxies, which is difficult to reconcile with evidence for an ongoing strong tidal interaction \citep[e.g.][]{Pearson2018,Liu2023}.
Owing to this inconsistency, several studies have instead adopted a pair-averaged distance of 7.14 Mpc \citep{Theureau2007}. Furthermore, a recent reanalysis of the colour–magnitude diagrams by \citet{Karachentsev2024} indicates that the TRGB location has been determined unreliably for NGC 4490/85.
Hence, in this work, we adopt a distance of 7.8 Mpc for NGC 4490 based on the TF relation when computing luminosities in order to maintain direct consistency with previous X-ray studies \citep{Roberts2002, Fridriksson2008, Gladstone2009, Yoshida2010a, Avdan2019a} of ULXs in this system.

\begin{figure*}
\centering

\includegraphics[width=3.3in]{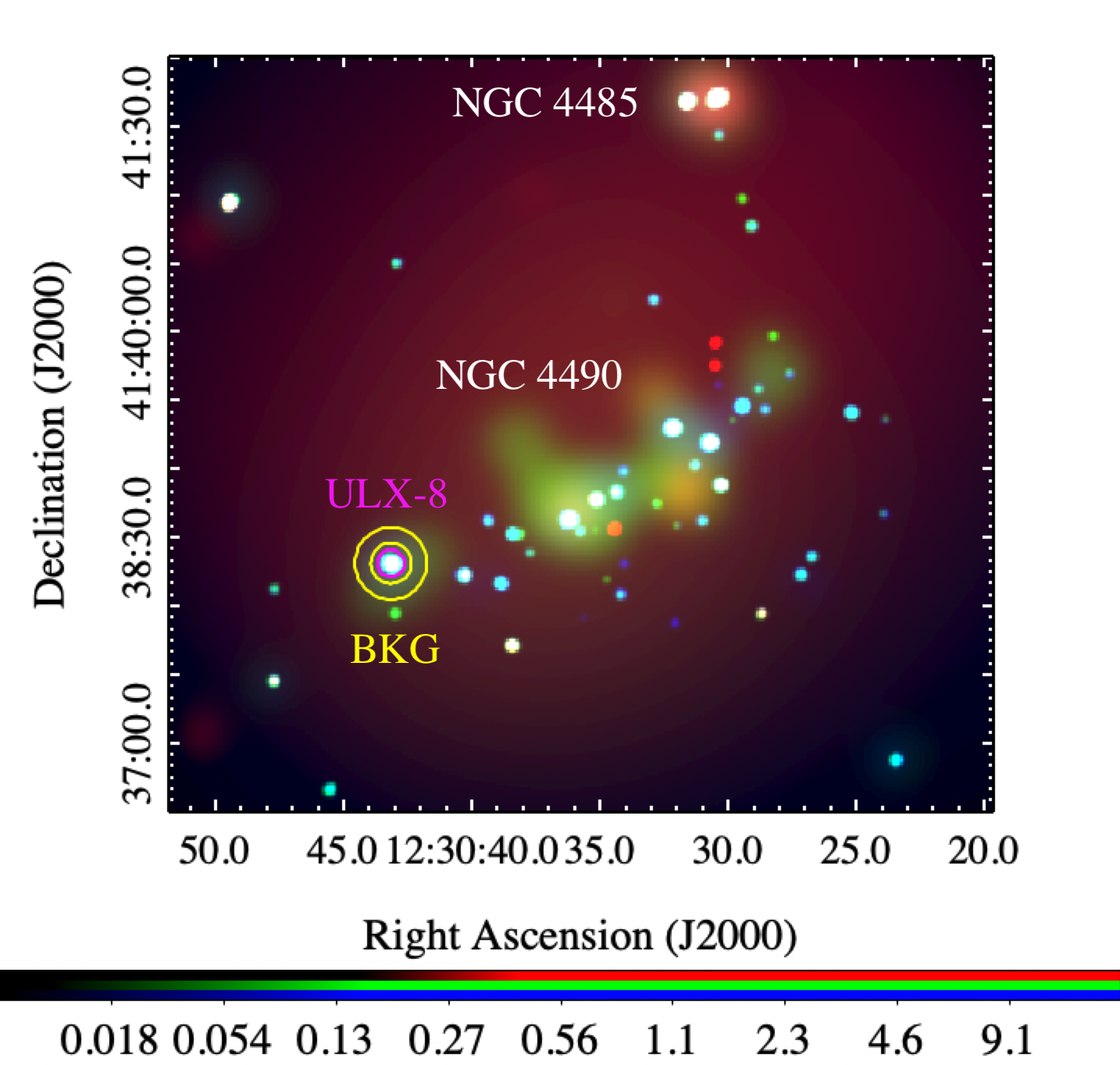} \hspace{0.3in}
\includegraphics[width=3.3in]{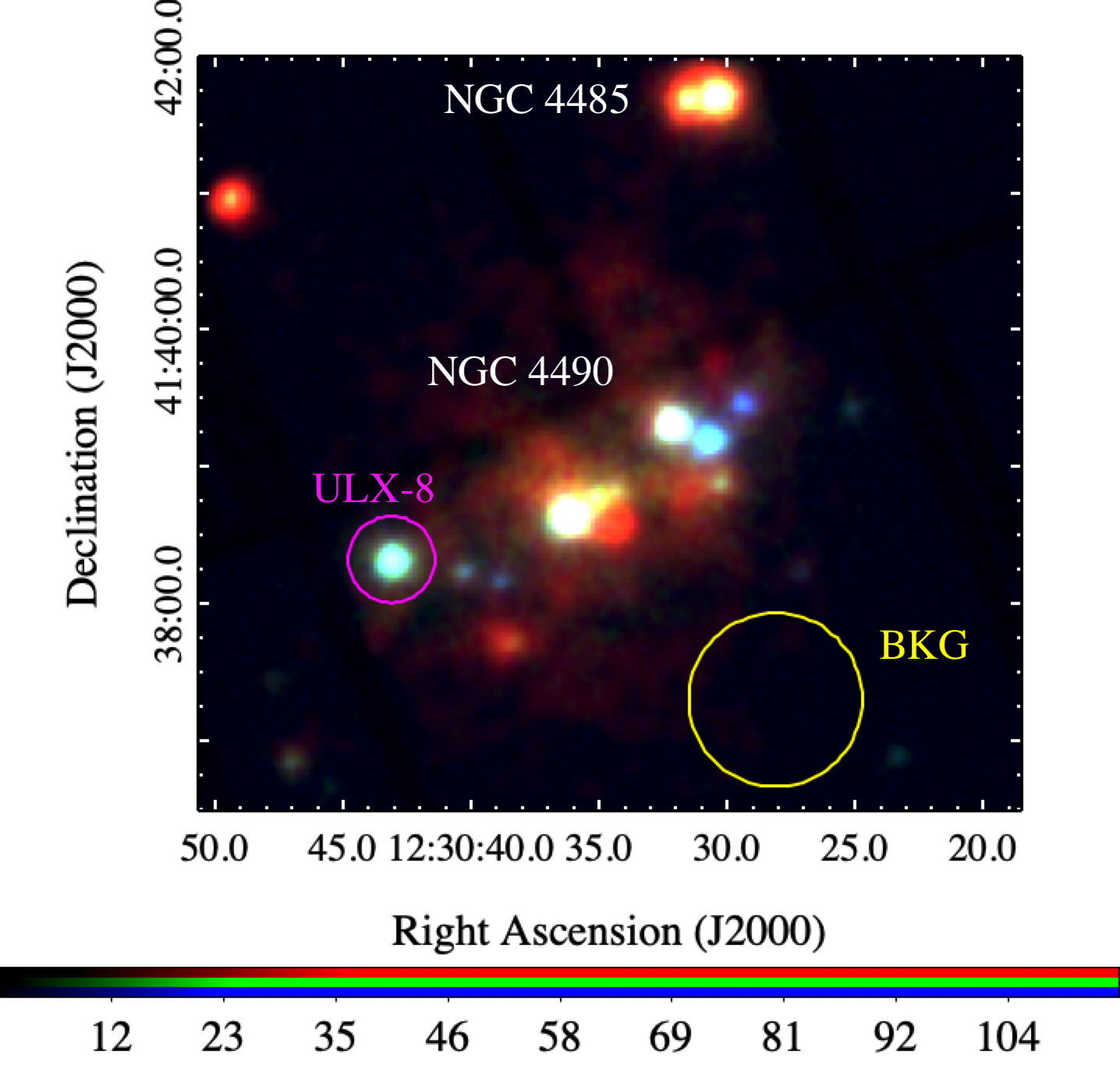}

\caption{False-colour X-ray images of NGC~4490/85 created from smoothed and merged observations. Representative source and background regions are marked in magenta and yellow colours, respectively. Red corresponds to emission in the 0.3--0.65 keV energy band, green to the 0.65--1.5 keV band and blue to the 1.5--6.0 keV band. The color scales are shown below each image. \textit{Left}: Image created from \textit{Chandra} observations. \textit{Right}: Image created from EPIC-pn and MOS1/2 observations with \textit{XMM-Newton}.}

\label{fig:NGC-4490}
\end{figure*}

This ongoing interaction of NGC 4490/85 is known to trigger active star formation, at a constant rate of $\sim 4.7$\,M$_{\sun}$\,yr$^{-1}$ \citep{Clemens1999, Clemens2002}, making this system an ideal laboratory for studying ULX populations. This interaction-driven star formation is likely responsible for the development of an extensive neutral hydrogen (HI) envelope encompassing the galaxy.
NGC 4490 has been observed multiple times in X-rays, beginning with \textit{ROSAT} \citep{Read1997, Roberts2000}. \citet{Roberts2002} studied the C1 \textit{Chandra} observation (see Table~\ref{ObservationTable}) to resolve X-ray sources in the galaxy, identifying 6 ULXs and a total of 36 X-ray sources. \citet{richings_hot_2010} examined the diffuse hot interstellar medium of NGC~4490 and catalogued the discrete X-ray source population using the first three \textit{Chandra} observations (C1--C3). This diffuse soft X-ray emission is also visible in the merged images presented in Figure~\ref{fig:NGC-4490}.
We adopt the ULX nomenclature of \citet{Fridriksson2008}, who catalogued 8 ULXs (ULX-1 through ULX-8) and 38 discrete X-ray sources based on C1-C3 and \textit{ROSAT}. 

Among NGC 4490's known ULXs, ULX-8 (CXOU~J123043.1+413818; \citealt{Fridriksson2008}) is a particularly intriguing candidate for investigating long-term timing and spectral variability. ULX-8 stands out as one of the brightest, along with ULX-3, ULX-4 and ULX-6. Notably, the position of ULX-6 is consistent with the nucleus of NGC 4490, as identified in recent optical and infrared studies by \citet{Lawrence2020}.
ULX-1, 3 and 4 lie in close proximity, complicating their analysis using \textit{XMM-Newton} due to source contamination. Extraction regions for these sources would need to be smaller than 9~arcsec to avoid overlap, limiting the available counts and affecting spectral fidelity. In contrast, ULX-8 is relatively isolated, allowing source extraction from circular regions as large as 19 arcsec in \textit{XMM-Newton}, thereby ensuring high-quality data suitable for long-term spectral and timing analysis. Previous studies of ULX-8 were based on limited datasets: \citet{Gladstone2009} and \citet{Yoshida2010a} analysed the \textit{Chandra} C1--3 and \textit{XMM-Newton} XM1 observations (see Table~\ref{ObservationTable}). \citet{Avdan2019a} identified the optical counterparts of ULXs in the NGC~4490/4485 system using archival \textit{HST} imaging, combined with \textit{Chandra} C1--3 and \textit{XMM-Newton} XM1--4 observations, and examined their long-term X-ray variability. Using astrometry, they determined the X-ray position for the source, with coordinates RA = 12:30:43.180 and Dec = +41:38:18.72. More recently, \citet{Earnshaw2025} presented results from a \textit{Chandra} Large Program that monitored 36 ULXs across three nearby galaxies, including NGC 4490, to study long-term variability behaviours.

We present the most extensive investigation of NGC 4490 ULX-8 to date, drawing on the full suite of archival \textit{Chandra} and \textit{XMM-Newton} observations together with 19 \textit{Swift}-XRT datasets. By unifying both analysed epochs (XM1–4, C1–3, C5–14) and several unexplored observations (XM5, XM6, C4, S1–S19), we are able to track the temporal and spectral evolution of ULX-8 across short (intra-observation), intermediate (weeks–months) and long (years) timescales. Our analysis aims to:
\begin{enumerate}
    \item Track variability in the unabsorbed X-ray luminosity of the source across two decades.
    \item Examine potential correlations among key spectral fitting parameters using statistical tools.
    \item Assess the physical implications of our findings for the accretion regime and the nature of the compact object.
\end{enumerate}

\section{Observations and Data Reduction}
\label{Obs}

\begin{table}
\setlength{\tabcolsep}{3pt}
\centering
\caption{Summary of observations for NGC~4490 ULX-8. Exposure times (in ks), after flare removal, are listed for EPIC-pn/MOS1/MOS2 detectors for \textit{XMM-Newton}, ACIS-S for \textit{Chandra} and the X-ray Telescope (XRT) for \textit{Swift}.}
\label{ObservationTable}
\begin{tabular}{@{}llccc@{}}
\toprule
\textbf{Mission} & \textbf{Epoch ID} & \textbf{Obs. ID} & \textbf{Date} & \textbf{Exposure (ks)} \\ \midrule
\textit{Chandra} & C1  & 1579       & 03/11/2000 & 19.52   \\
                 & C2  & 4725       & 29/07/2004 & 38.47   \\
                 & C3  & 4726       & 20/11/2004 & 38.74   \\
                 & C4  & 20999      & 06/11/2018 & 14.79   \\
                 & C5  & 23482      & 27/11/2020 & 28.68   \\
                 & C6  & 23483      & 27/12/2020 & 29.47   \\
                 & C7  & 23484      & 24/01/2021 & 29.37   \\
                 & C8  & 23485      & 21/02/2021 & 29.87   \\
                 & C9  & 23486      & 19/03/2021 & 29.66   \\
                 & C10 & 23487      & 18/04/2021 & 29.57   \\
                 & C11 & 23488      & 15/05/2021 & 29.47   \\
                 & C12 & 23489      & 12/06/2021 & 29.59   \\
                 & C13 & 23490      & 08/07/2021 & 29.38   \\
                 & C14 & 23491      & 06/08/2021 & 29.56   \\ \midrule
\textit{XMM-Newton} & XM1 & 0112280201 & 27/05/2002 & 10.21/15.37/15.38 \\
                    & XM2 & 0556300101 & 19/05/2008 & 15.25/18.64/18.64 \\
                    & XM3 & 0556300201 & 22/06/2008 & 23.66/28.72/28.73 \\
                    & XM4 & 0762240201 & 12/06/2015 & 7.59/10.04/10.04 \\
                    & XM5 & 0824450201 & 29/12/2018 & 36.49/55.15/55.23 \\
                    & XM6 & 0891020301 & 30/05/2022 & 22.71/28.14/17.67 \\ \midrule
\textit{Swift}-\textsc{XRT} & S1  & 00031155001 & 04/03/2008 & 4.93  \\
                   & S2  & 00031155002 & 07/03/2008 & 4.23  \\
                   & S3  & 00031155004 & 12/03/2008 & 3.13  \\
                   & S4  & 00031155006 & 18/03/2008 & 4.28  \\
                   & S5  & 00031155010 & 21/03/2008 & 2.04  \\
                   & S6  & 00031155016 & 07/04/2008 & 2.21  \\
                   & S7  & 00031155017 & 11/04/2008 & 1.88  \\
                   & S8  & 00031155019 & 17/04/2008 & 2.60  \\
                   & S9  & 00031155020 & 18/04/2008 & 2.36  \\
                   & S10 & 00031155021 & 22/04/2008 & 5.85  \\
                   & S11 & 00031155022 & 25/04/2008 & 4.17  \\
                   & S12 & 00031155023 & 26/04/2008 & 4.35  \\
                   & S13 & 00037784001 & 22/11/2008 & 13.20 \\
                   & S14 & 00085424002 & 30/11/2014 & 2.48  \\
                   & S15 & 00085424013 & 31/05/2015 & 2.44  \\
                   & S16 & 00085424014 & 12/08/2015 & 3.84  \\
                   & S17 & 00089252002 & 14/06/2022 & 1.49  \\
                   & S18 & 00015121005 & 01/12/2022 & 1.52  \\
                   & S19 & 00015121012 & 04/05/2024 & 1.66  \\ \bottomrule
\end{tabular}
\end{table}

In this work, we study 6 \textit{XMM-Newton} \citep{Jansen2001}, 14 \textit{Chandra} \citep{Weisskopf1999} and 19 \textit{Swift}-\textsc{XRT} \citep{Burrows2005} archival observations of NGC 4490 ULX-8. In this paper, we label observations as XM1-6, C1-C14 and S1-S19 (Table~\ref{ObservationTable}). We have used all available \textit{Chandra} and \textit{XMM-Newton} observations, except for \textit{XMM-Newton} Obs. ID 0891020501 and \textit{Chandra} Obs. ID 28115, due to their low post-reduction exposure times. For \textit{Swift}-\textsc{XRT} observations, we have filtered the archival dataset for a detection significance above 3$\sigma$, leading to 19 observations (Table~\ref{ObservationTable}) for further analysis. The corresponding \textit{Swift}-\textsc{XRT} spectral products had sufficient counts, based on count rates and exposure times (see Table~\ref{Swift-Parameter-table}), to justify the use of C-statistics.
Some of these \textit{Swift}-\textsc{XRT} observations, as well as the \textit{Chandra} observation C4, were originally targeted at SN2008ax, a supernova in NGC 4490. However, ULX-8 is detected in these fields and is therefore included in our analysis.

\subsection{\textit{XMM-Newton}}
\label{sec:xmm-newton} 
We used \texttt{Scientific Analysis Software}\footnote{\url{https://www.cosmos.esa.int/web/XMM-Newton}} (SAS) v.~21.0.0 to process all the observations. Standard \texttt{emproc} and \texttt{epproc} tasks were used to reprocess the Observation Data Files (ODFs) for the MOS and pn detectors, respectively.
A filtering constraint of \texttt{FLAG==0} was used for the EPIC cameras. Data were reduced by selecting single to quadruple events (\texttt{PATTERN~$\leq$~12}) and energies within 0.2--12~keV for the MOS detector and by selecting single and double events (\texttt{PATTERN~$\leq$~4}) and energies within 0.2--15~keV for pn. To identify intervals of flaring particle background, we generated single-event (\texttt{PATTERN == 0}), high-energy light curves\footnote{\url{https://www.cosmos.esa.int/web/xmm-newton/sas-thread-epic-filterbackground}} using the \texttt{evselect} task, choosing energies above 10~keV for MOS and between 10--12~keV for pn detectors. Subsequently, we created a Good Time Interval (GTI) file, which was used to filter the event files using the same task. While the filtering reduced the effective exposure times for most observations by some margins, it doesn't impact the reliability of our spectral and timing analyses. We extracted source products from circular regions of radius 19~arcsec centred at ULX-8. The background region was chosen from a circle of radius 38~arcsec on the same charge-coupled device (CCD) as the source region (Figure~\ref{fig:NGC-4490}, \textit{Right} panel). A barycentric correction for timing analysis was performed using the \texttt{barycen} task with the known coordinates of the source. We used \texttt{evselect} to create source and background products. The \texttt{epiclccorr} task was used to generate background-subtracted light curves, applying various absolute and relative corrections. The redistribution matrix file (RMF) and ancillary response file (ARF) were generated using the \texttt{rmfgen} and \texttt{arfgen} tasks, respectively. These were then grouped with the source and background spectra using the \texttt{specgroup} task with a minimum of 20 counts per energy bin to ensure adequate statistics for spectral fitting.

\subsection{\textit{Chandra}}
The \textit{Chandra} data were analysed using \texttt{CIAO}\footnote{\url{https://cxc.cfa.harvard.edu/ciao/}} v.~4.15, employing the calibration files (\texttt{CALDB}\footnote{\url{https://cxc.cfa.harvard.edu/caldb/}} v.~4.10.4). All the observations were performed with the Advanced CCD Imaging Spectrometer (ACIS) with the aim point placed on the S3 chip of the ACIS-S array. The raw event 1 files were reprocessed and \texttt{VFAINT} mode was used for observations C2--C4. We restricted the data to the energy range 0.3--10~keV and checked for flares, which did not affect any of the observations. We also confirmed that the ACIS data had source counts below the temperature limit of 164.15~K, in accordance with the `Removing Warm ACIS Data' thread\footnote{\url{https://cxc.cfa.harvard.edu/ciao/threads/acisfptemp/}}. 
ULX-8 is located off-axis in most \textit{Chandra} pointings, where the broader point spread function (PSF) required the use of elliptical extraction regions for source products. These were centred on the source coordinates, with semi-major and semi-minor axes of 6 and 4.5 pixels, respectively. Background spectra were extracted from nearby source-free circular regions with 20-pixel diameter. In contrast, observations C1, C4, C13 and C14 captured the source closer to the optical axis. For these, a circular region of 5 pixels in diameter was used to extract source counts, encircling more than 90 \% of the PSF, while background events were obtained from an annulus with inner and outer diameter of 8 and 16 pixels, respectively (represented in Figure~\ref{fig:NGC-4490}, \textit{Left} panel). For timing analysis, the CIAO tool \texttt{axbary} was used for barycentric correction. The \texttt{dmextract} and \texttt{specextract} tools were used to extract the source light curves and spectra, respectively. The extracted spectra were grouped to a minimum of 20 counts per bin.

\subsection{\textit{Swift}-\textsc{XRT}}

The \textit{Swift}-\textsc{XRT} observations were performed in photon counting (PC) mode. We analysed all available \textit{Swift}-\textsc{XRT} \citep{Burrows2005} data using the standard online processing tools provided by the UK Swift Science Data Centre \citep{Evans2007, Evans2009}. Source positions were determined using the simple centroid method with the \texttt{SIMBAD} coordinates of ULX-8 as input, adopting a 1$\arcsec$ positional uncertainty. We selected only those observations with a detection significance more than 3 $\sigma$, ensuring reliable source detection while minimising contamination from background noise. We also investigated possible contamination from the nearby supernova SN2008ax. Using \texttt{XIMAGE} we examined the region around the SN's \texttt{SIMBAD} position (RA = 187.6700°, Dec = 41.6378°) but did not detect any significant X-ray emission at the SN location in the \textit{Swift-XRT} images. The angular separation between the ULX and the SN position is $\sim$ 28 $\arcsec$, above the \textit{Swift-XRT} half-power diameter of $\sim$ 18 $\arcsec$, confirming that the ULX extraction is unaffected by the SN.

\section{Analysis}
\subsection{Timing Analysis}
\label{timing}

\begin{figure}
 \includegraphics[width=\columnwidth]{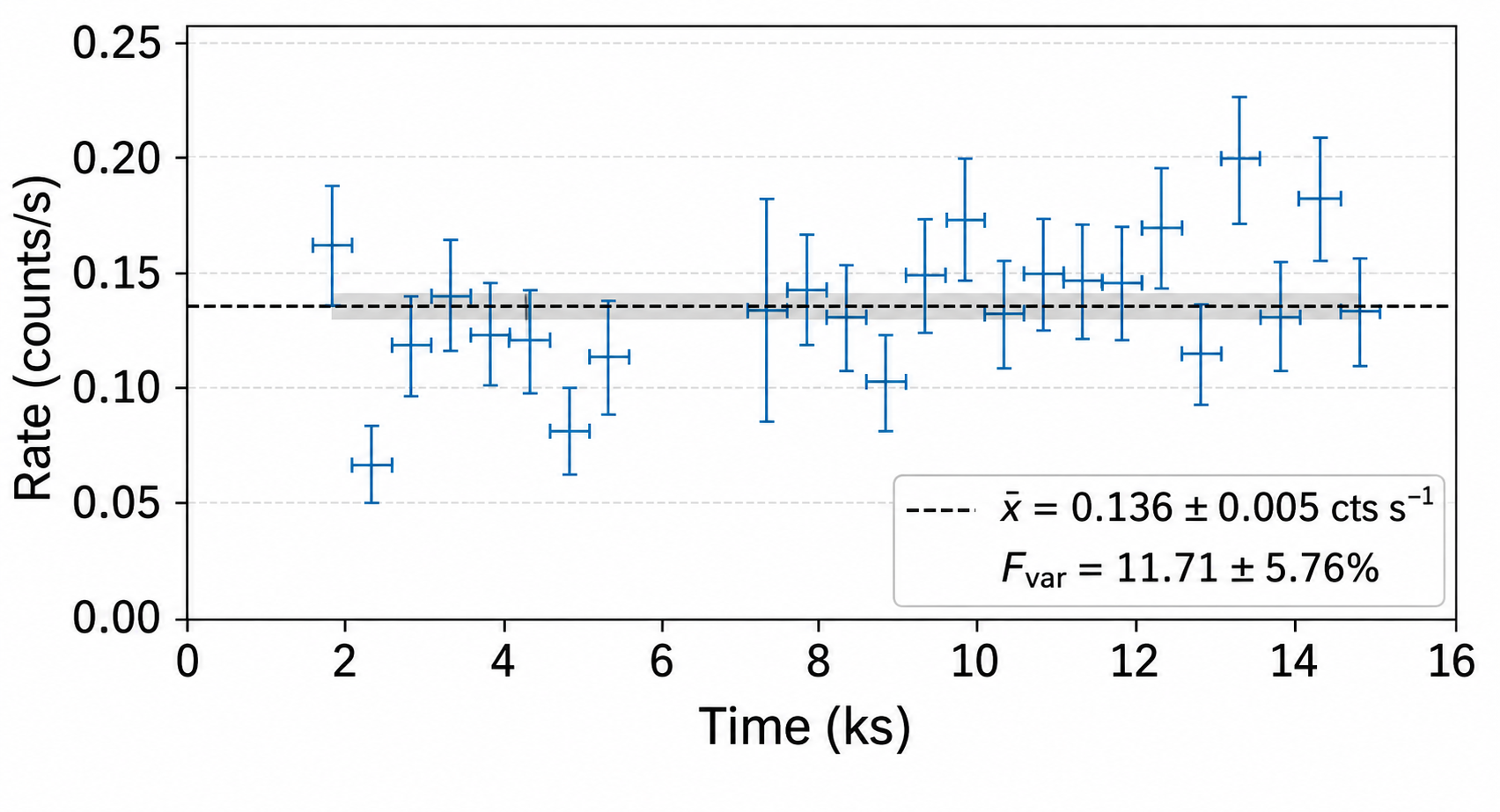}
 \caption{Background subtracted light curve with 500s bin-time from the \textit{XMM-Newton} EPIC-pn module during the XM1 observation. The dashed line marks the mean count rate, while the shaded region indicates its uncertainty.}
 \label{fig:lightcurve}
\end{figure}

To probe the time series, we generated light curves of all observations. Data from the pn module of \textit{XMM-Newton} has been used for this analysis. We quantify the short-term variability within individual observations using the fractional root-mean-square (rms) variability amplitude, $F_{\rm var}$, as defined by \citet{Vaughan2003}. No statistically significant short-term variability is detected in the majority of the light curves. However, mild variability was observed in some observations, for example XM1 and XM4, with $F_{\rm var}$ of 11.71 $\pm$ 5.76 \% and 11.74 $\pm$ 7.16 \% respectively, for $500~\mathrm{s}$ binned light curves. A representative light curve for XM1 is plotted in Figure~\ref{fig:lightcurve}.

To search for possible periodicities in the source light curves, we used the \texttt{ftool} task \texttt{POWSPEC}\footnote{\url{https://heasarc.gsfc.nasa.gov/lheasoft/help/powspec.html}} v1.0 to generate the power spectra. No significant periodic signal was detected in any observation. To confirm this, we employed \texttt{HENDRICS}\footnote{\url{https://hendrics.stingray.science/en/latest/}} v8.0.2 \citep{2018ascl.soft05019B}, a set of \texttt{STINGRAY}-based command-line scripts. Analysis with the \texttt{HENaccelsearch} and \texttt{HENzsearch} tasks of \texttt{HENDRICS} also revealed no pulsations in the light curves.

\subsection{Hardness–intensity diagrams}
\label{HID}

\begin{figure}
 \includegraphics[width=\columnwidth]{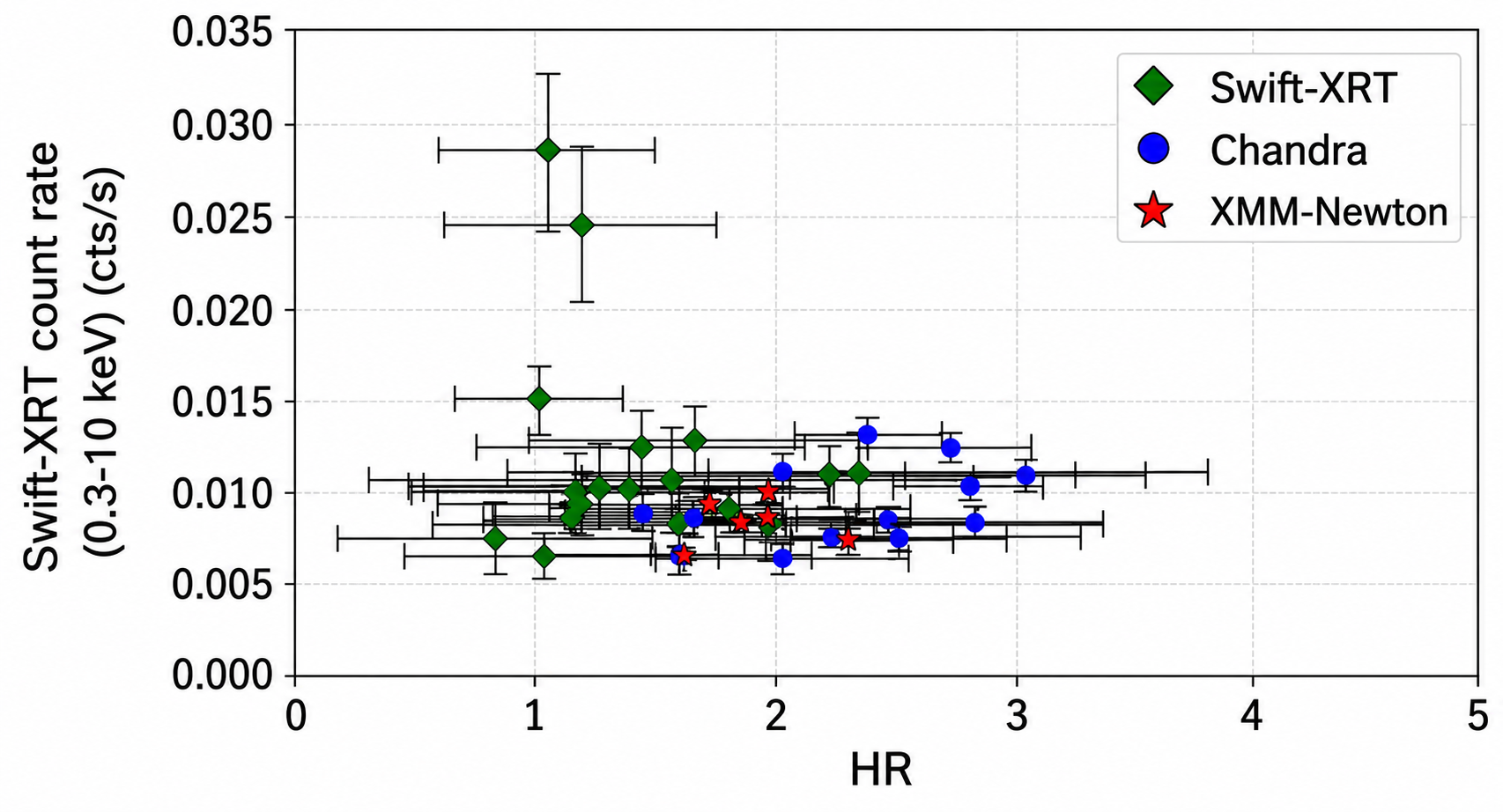}
  \caption{The hardness-intensity diagram: Total 0.3--10 keV count rate vs hardness ratio (HR), using observations from multiple instruments. Data points from \textit{Swift}–XRT are shown in green, \textit{Chandra} data in blue and \textit{XMM-Newton} data in red.}
 \label{fig:HID}
\end{figure}

Hardness–intensity diagrams (HIDs) have long been employed to track spectral state transitions in both ULXs and X-ray binaries (XRBs). In Galactic XRBs, HIDs plotting X-ray hardness versus intensity reveal characteristic \textit{q‑shaped} tracks that correspond to transitions among canonical states such as low/hard, high/soft and intermediate, as shown by GX 339–4’s 2002/2003 outburst \citep{Belloni2010}. In ULXs, evolving patterns on the HID plane have been increasingly recognized: for instance, \citet{Gurpide2021a} demonstrated recurrent spectral cycles between hard-ultraluminous and soft-ultraluminous regimes in sources like Holmberg II X‑1 and NGC 5204 X‑1 via HID tracking. 

In Figure~\ref{fig:HID}, we present the HID of NGC 4490 ULX-8, where the hardness ratio (HR) is plotted as a function of the total 0.3–10 keV count rate converted to the \textit{Swift}-\textsc{XRT} reference band. HR is computed as the ratio of the count rates in the 1.5--10~keV band to that in the 0.3--1.5~keV band. We find the HR of NGC 4490 ULX-8 to be in the range of 0.84-3.04 with typical measurement uncertainties ranging from $\sim$0.15 to 1.45. To account for instrument differences, we have used the standard online \texttt{WebPIMMS}\footnote{\url{https://heasarc.gsfc.nasa.gov/cgi-bin/Tools/w3pimms/w3pimms.pl}} \citep{Mukai1993} tool to convert from \textit{Chandra} and \textit{XMM-Newton} count rates to \textsc{XRT} assuming a redshift of $z = 0.0018$ \citep{Strauss1992a}.

We find that the most recent observations in our dataset, S18 and S19, taken on 2022 December 1 and 2024 May 4 respectively exhibit significantly higher count rates compared to earlier epochs. However, no distinct hard or soft spectral states could be identified from the HID.The observed increase in count rate may therefore reflect changes in the accretion rate rather than a canonical state transition \citep{Mondal2021}.

\subsection{Spectral Analysis}
\label{SA}

\begin{table*}
	\centering
    \renewcommand{\arraystretch}{1.2}
    \setlength{\tabcolsep}{5.5pt}
	\caption{Best-fitting spectral parameters for NGC 4490 ULX-8 across different \textit{Chandra} and \textit{XMM-Newton} observations, obtained using the \texttt{tbabs*tbabs*powerlaw} and \texttt{tbabs*tbabs*diskbb} models.}
	\label{ParameterTable}
	\begin{tabular}{@{}cccccccccc@{}}
\toprule Epoch & Model  & $N_{\mathrm{H,int}}^{a}$ & $\Gamma^{b}$                   & \textit{$T_{in}^{c}$} & Disk Norm (\textit{N})$^{d}$          & Unabsorbed $L_{X}^{e}$          & Absorbed $L_{X}^{e}$           & ${\chi^{2}}$/d.o.f.$^{f}$ &  Simftest P-value$^{g}$ \\ 
      &        & $[10^{22}cm^{-2}]$                                &                        & [keV]      &  [$10^{-2}$]              & [$10^{39}\ \mathrm{erg\ s^{-1}}$]                        & [$10^{39}\ \mathrm{erg\ s^{-1}}$]                         &              \\\midrule
C1    & PL     & 1.18$^{+0.20}_{-0.18}$            & 2.16$^{+0.21}_{-0.20}$  & --  & --                   & 6.73$^{+1.01}_{-0.76}$       & 3.53$^{+0.29}_{-0.28}$   & 47.58/41    & 0.03 \\
      & DISKBB & 0.73$^{+0.13}_{-0.12}$          & --                     & 1.22$^{+0.14}_{-0.12}$ & 1.40$^{+0.76}_{-0.50}$ &
      4.39$^{+0.33}_{-0.32}$    & 3.21$^{+0.26}_{-0.25}$   & \textbf{39.58/41 }   & 0.77 \\ \midrule

C2    & PL     & 1.26$^{+0.17}_{-0.15}$          & 2.67$^{+0.19}_{-0.18}$ & --  & --                   & 5.76$^{+1.02}_{-0.77}$       & 2.11$^{+0.14}_{-0.13}$      & 73.63/55  & $4\times 10^{-3*}$   \\
      & DISKBB & 0.67$\pm$0.10               & --                     & 0.97$^{+0.08}_{-0.07}$ & 2.43$^{+0.97}_{-0.69}$ &2.90$^{+0.20}_{-0.19}$    & 1.98$\pm$0.12       & \textbf{58.67/55}  & 0.56   \\ \midrule
C3    & PL     & 1.00$^{+0.15}_{-0.13}$          & 2.06$^{+0.15}_{-0.14}$ & --  & --                   & 5.78$^{+0.56}_{-0.47}$    & 3.34$^{+0.19}_{-0.18}$   & 85.92/73   & 0.05  \\
      & DISKBB & 0.52$^{+0.10}_{-0.09}$           & --                     & 1.37$^{+0.13}_{-0.11}$ & 0.79$^{+0.32}_{-0.23}$ &3.93$\pm$0.21    & 3.13$\pm$0.18       & \textbf{84.72/73}    & 0.05 \\ \midrule
C4    & PL     & 1.13$^{+0.69}_{-0.60}$           & 2.36$^{+0.51}_{-0.47}$ & -- & --                    & 4.85$^{+2.84}_{-1.33}$     & 2.28$^{+0.34}_{-0.31}$   & 20.46/15   & 0.12  \\
      & DISKBB & 0.42$^{+0.47}_{-0.41}$          & --                     & 1.14$^{+0.31}_{-0.21}$ & 
      1.14$^{+1.74}_{-0.72}$ &2.66$^{+0.47}_{-0.38}$   & 2.09$^{+0.34}_{-0.27}$   & \textbf{17.95/15}   & 0.54  \\ \midrule
C5    & PL     & 0.61$^{+0.51}_{-0.45}$          & 1.92$^{+0.37}_{-0.34}$ & --  & --                   & 2.90$^{+0.83}_{-0.50}$    & 1.98$^{+0.26}_{-0.22}$   & \textbf{12.06/21}   & 0.51 \\
      & DISKBB & $<0.29$                    & --                     & $1.59^{+0.26}_{-0.20}$   &
       $0.23^{+0.14}_{-0.09}$ & 1.99$\pm$0.18        & 1.97$\pm$0.18       & 13.36/21 & 0.24    \\ \midrule
C6    & PL     & $0.33^{+0.46}_{-0.33}$           & $1.74^{+0.37}_{-0.33}$ & -- & --                    & $2.52^{+0.55}_{-0.35}$   & $2.02^{+0.35}_{-0.25}$   & \textbf{18.80/21} & 0.45   \\
      & DISKBB & $<0.15$                   & --                     & 1.62$^{+0.31}_{-0.23}$ & 
      0.20$^{+0.16}_{-0.09}$ &1.92$\pm$0.18   & 1.90$\pm$0.18   & 24.27/21 & 0.05   \\ \midrule
C7    & PL     & 1.24$^{+0.45}_{-0.40}$           & 2.49$^{+0.33}_{-0.30}$  & --  & --                   & 5.29$^{+1.95}_{-1.17}$     & 2.21$^{+0.20}_{-0.19}$   & \textbf{15.51/28} & 0.75    \\
      & DISKBB & 0.34$^{+0.29}_{-0.26}$          & --                     & 1.19$^{+0.19}_{-0.16}$ & 
      0.91$^{+0.79}_{-0.43}$ &2.60$^{+0.28}_{-0.25}$   & 2.14$^{+0.21}_{-0.19}$   & 18.98/28 & 0.17    \\ \midrule
C8    & PL     & 0.87$^{+0.29}_{-0.27}$          & 2.28$^{+0.24}_{-0.23}$ & -- & --                    & 6.12$^{+1.30}_{-0.91}$    & 3.26$^{+0.25}_{-0.24}$  & \textbf{39.50/40} & 0.81    \\
      & DISKBB & 0.15$^{+0.20}_{-0.15}$           & --                     & 1.30$^{+0.19}_{-0.15}$  &
      0.88$^{+0.61}_{-0.37}$ &3.51$^{+0.27}_{-0.26}$   & 3.17$^{+0.29}_{-0.26}$   & 51.73/40  & $4 \times 10^{-3}$$^*$  \\ \midrule
C9    & PL     & 1.19$^{+0.29}_{-0.26}$           & 2.49$^{+0.22}_{-0.21}$ & -- & --                    & 9.00$^{+2.04}_{-1.43}$    & 3.83$\pm$0.24       & 42.27/50   & 0.10  \\
      & DISKBB & 0.42$^{+0.18}_{-0.16}$          & --                     & 1.15$^{+0.11}_{-0.10}$  & 
      1.92$^{+0.98}_{-0.65}$ &4.64$^{+0.35}_{-0.32}$   & 3.66$^{+0.24}_{-0.23}$   & \textbf{37.88/50}  & 0.62   \\ \midrule
C10   & PL     & 0.49$^{+0.37}_{-0.33}$          & 2.11$^{+0.28}_{-0.26}$ & -- & --                    & 3.70$^{+0.88}_{-0.59}$   & 2.47$^{+0.28}_{-0.23}$   & \textbf{28.15/30} & 0.22    \\
      & DISKBB & $<0.13$                    & --                     & 1.31$^{+0.14}_{-0.12}$ & 
      0.58$^{+0.26}_{-0.19}$ &2.45$\pm$0.17   & 2.42$\pm$0.17   & 28.60/30  & 0.19   \\ \midrule
C11   & PL     & 0.68$^{+0.44}_{-0.39}$           & 2.25$^{+0.33}_{-0.30}$  & -- & --                    & 4.15$^{+1.28}_{-0.79}$    & 2.41$^{+0.28}_{-0.23}$   & 20.44/28   & 0.36    \\
      & DISKBB &  $<0.82$                   & --                     & 1.26$^{+0.14}_{-0.12}$ & 
      0.68$^{+0.31}_{-0.22}$ &2.45$^{+0.18}_{-0.17}$   & 2.41$^{+0.19}_{-0.16}$   & \textbf{19.67/28}  & 0.51    \\ \midrule
C12   & PL     & 1.36$^{+0.34}_{-0.3}$           & 2.53$^{+0.22}_{-0.21}$ & --  & --                   & 9.39$^{+2.36}_{-1.65}$     & 3.70$^{+0.23}_{-0.22}$   & 49.53/50   & $4 \times     
10^{-3}$$^*$  \\
      & DISKBB & 0.43$^{+0.22}_{-0.19}$          & --                     & 1.21$^{+0.12}_{-0.11}$ & 
      1.55$^{+0.79}_{-0.53}$ &4.60$^{+0.37}_{-0.34}$   & 3.65$\pm$0.23       & \textbf{35.98/50} & 0.91   \\ \midrule
C13   & PL     & 1.27$^{+0.58}_{-0.50}$           & 2.38$^{+0.36}_{-0.33}$ & -- & --                    & 5.78$^{+2.45}_{-1.37}$   & 2.59$^{+0.24}_{-0.23}$   & 25.48/37   & 0.02 \\
      & DISKBB & 0.41$^{+0.38}_{-0.33}$          & --                     & 1.23$^{+0.21}_{-0.17}$ & 
      0.96$^{+0.89}_{-0.47}$ &3.09$^{+0.38}_{-0.35}$   & 2.48$^{+0.26}_{-0.23}$   & \textbf{18.84/37} & 0.27     \\ \midrule
C14   & PL     & 1.35$^{+0.47}_{-0.41}$          & 2.65$^{+0.3}_{-0.27}$  & --  & --                   & 6.75$^{+2.61}_{-1.59}$     & 2.44$^{+0.19}_{-0.18}$   & 41.13/34  & $6 \times  10^{-3*}$  \\
      & DISKBB & 0.34$^{+0.29}_{-0.26}$          & --                     & 1.14$^{+0.15}_{-0.13}$ & 
      1.26$^{+0.91}_{-0.52}$ &3.01$^{+0.34}_{-0.30}$   & 2.45$^{+0.22}_{-0.20}$   & \textbf{30.45/34} & 0.35     \\ \midrule \midrule
XM1    & PL     & 0.93$^{+0.15}_{-0.13}$          & 2.56$^{+0.19}_{-0.17}$ & --  & --                   & 5.66$^{+0.97}_{-0.76}$    & 2.50$\pm$0.21       & 69.07/66    &  0.02 \\
      & DISKBB & 0.41$^{+0.09}_{-0.08}$          & --                     & 0.99$^{+0.09}_{-0.08}$ & 
      2.28$^{+1.01}_{-0.69}$
      &3.04$^{+0.27}_{-0.26}$    & 2.31$\pm$0.19       & \textbf{61.44/66} & 0.26    \\ \midrule      
XM2    & PL     & 0.71$^{+0.17}_{-0.14}$          & 2.22$^{+0.18}_{-0.17}$ & --  & --                   & 3.83$^{+0.63}_{-0.51}$    & 2.22$^{+0.21}_{-0.20}$   & 96.81/62 & 0.00$^*$   \\
      & DISKBB & 0.28$^{+0.10}_{-0.08}$           & --                     & 1.21$^{+0.11}_{-0.10}$  & 
      0.86$^{+0.38}_{-0.26}$ &2.60$^{+0.26}_{-0.25}$    & 2.20$\pm$0.20       & \textbf{68.21/62}   & 0.34  \\ \midrule
XM3    & PL     & 0.77$^{+0.10}_{-0.09}$           & 2.18$\pm$0.13 & -- & --                    & 5.05$^{+0.46}_{-0.40}$    & 2.94$^{+0.19}_{-0.20}$   & 225.36/213  & 0.00$^{*}$ \\
      & DISKBB & 0.39$\pm$0.06              & --                     & 1.16$^{+0.09}_{-0.08}$ & 1.33$^{+0.44}_{-0.33}$ &3.38$\pm$0.21       & 2.71$^{+0.19}_{-0.18}$   & \textbf{191.03/213} & 0.50  \\ \midrule
XM4    & PL     & 0.91$^{+0.18}_{-0.16}$          & 2.37$^{+0.21}_{-0.19}$ & --  & --                   & 4.91$^{+0.89}_{-0.68}$   & 2.45$^{+0.23}_{-0.22}$   & 60.29/49   & 0.02  \\
      & DISKBB & 0.43$^{+0.10}_{-0.09}$           & --                     & 1.12$^{+0.11}_{-0.10}$  & 
      1.32$^{+0.66}_{-0.43}$ &2.94$^{+0.28}_{-0.27}$   & 2.30$^{+0.21}_{-0.20}$       & \textbf{51.95/49}   & 0.68  \\ \midrule
XM5    & PL     & 0.87$\pm$0.07              & 2.55$^{+0.09}_{-0.08}$ & -- & --                    & 3.83$^{+0.32}_{-0.28}$   & 1.75$\pm$0.07 & 251.87/166   & 0.00$^*$\\
      & DISKBB & 0.37$\pm$0.04              & --                     & 1.06$\pm$0.04 & 
      1.27$^{+0.24}_{-0.20}$ & 
      2.19$^{+0.10}_{-0.09}$ & 1.73$\pm$0.07 & \textbf{154.28/166} & 0.84  \\ \midrule      
XM6    & PL     & 0.74$^{+0.08}_{-0.07}$          & 2.20$\pm$0.10    & -- & --                    & 4.17$^{+0.31}_{-0.28}$   & 2.41$\pm$0.12       & \textbf{112.49/123} & $3\times 10^{-3*}$   \\
      & DISKBB & 0.35$\pm$0.04              & --                     & 1.24$\pm$0.07 &
      0.84$^{+0.22}_{-0.17}$ &2.82$\pm$0.14       & 2.32$\pm$0.12       & 119.34/123 & $1\times 10^{-3}$$^*$\\ 
\bottomrule
\end{tabular}

\raggedright 
\textbf{Notes.} Parameter uncertainties are quoted at the 90\% confidence level. 
$^a$ Intrinsic hydrogen column density. 
$^b$ Photon index. 
$^c$ Inner disc temperature of the \texttt{DISKBB} component. 
$^d$ \texttt{DISKBB} normalisation values. 
$^e$ Luminosity calculated in the 0.5--8\,keV band. 
$^f$ $\chi^2$ and degrees of freedom; the lower reduced $\chi^2$ per observation is highlighted. 
$^g$ F-test statistic and corresponding probability with respect to \texttt{tbabs*tbabs(powerlaw+diskbb)}. 
$^*$ Probability value $< 0.01$.  \\
\end{table*}

\begin{table*}
   \centering
   \renewcommand{\arraystretch}{1.2}
   \setlength{\tabcolsep}{9pt}
    \caption{Best-fitting spectral parameters for NGC 4490 ULX-8 from different \textit{Swift}-\textsc{XRT} observations, obtained using the \texttt{tbabs*tbabs*powerlaw} model.}
\label{Swift-Parameter-table}
\begin{tabular}{@{}cccccccc@{}}
\toprule
Epoch & Count Rate$^a$ & Exposure Time & $N_{\mathrm{H,int}}^{b}$ & $\Gamma$$^c$ & Unabsorbed $L_{X}$$^d$ & Absorbed $L_{X}$$^d$ & $C\text{-stat}/d.o.f.$$^e$ \\
& $[10^{-2}\ \mathrm{cts\ s^{-1}}]$ & [ks] & $[10^{22}\ \mathrm{cm^{-2}}]$ & & [$10^{39}\ \mathrm{erg\ s^{-1}}$] & [$10^{39}\ \mathrm{erg\ s^{-1}}$] & \\ \hline
S1  & 1.475 $\pm$ 0.176 & 4.93 & 0.32$^{+0.18}_{-0.14}$ & 2.08$^{+0.38}_{-0.34}$ & 6.16$^{+1.31}_{-0.96}$ & 4.53$^{+0.81}_{-0.69}$ & 63.43/60 \\ \hline
S2  & 1.270 $\pm$ 0.178 & 4.23 & 0.49$^{+0.30}_{-0.23}$ & 1.84$^{+0.44}_{-0.40}$ & 6.67$^{+1.70}_{-1.14}$ & 4.89$^{+1.04}_{-0.86}$ & 48.71/48 \\ \hline
S3  & 1.220 $\pm$ 0.203 & 3.13 & 0.25$^{+0.24}_{-0.18}$ & 1.45$^{+0.43}_{-0.40}$ & 5.88$^{+1.25}_{-1.05}$ & 5.14$^{+1.28}_{-1.04}$ & 38.24/35 \\ \hline
S4  & 1.069 $\pm$ 0.162 & 4.28 & 0.51$^{+0.60}_{-0.40}$ & 1.35$^{+1.07}_{-0.91}$ & 7.94$^{+4.93}_{-1.99}$ & 6.59$^{+5.81}_{-2.75}$ & 31.82/35 \\ \hline
S5  & 0.736 $\pm$ 0.196 & 2.04 & 0.25$^{+0.40}_{-0.25}$ & 1.83$^{+1.12}_{-0.95}$ & 3.74$^{+1.92}_{-1.12}$ & 3.06$^{+2.43}_{-1.18}$ & 10.25/13 \\ \hline
S6  & 1.013 $\pm$ 0.226 & 2.21 & 0.75$^{+0.59}_{-0.41}$ & 2.34$^{+0.81}_{-0.69}$ & 6.82$^{+6.02}_{-2.14}$ & 3.67$^{+1.26}_{-0.93}$ & 27.69/21 \\ \hline
S7  & 0.837 $\pm$ 0.219 & 1.88 & 0.47 (fixed) & 1.50 $\pm$ 0.48 & 
4.65$^{+1.58}_{-1.20}$ & 3.77$^{+1.61}_{-1.15}$ & 15.79/14 \\ \hline
S8  & 0.989 $\pm$ 0.200 & 2.60 & 0.47 (fixed) & 2.10 $\pm$ 0.43 & 
5.78$^{+1.32}_{-1.07}$ & 3.87$^{+1.38}_{-0.98}$ & 20.48/22 \\ \hline
S9  & 1.009 $\pm$ 0.212 & 2.36 & 0.47 (fixed) & 2.00 $\pm$ 0.41 & 
5.45$^{+1.29}_{-1.08}$ & 3.81$^{+1.29}_{-0.99}$ & 14.88/22 \\ \hline
S10 & 0.896 $\pm$ 0.127 & 5.85 & 0.44$^{+0.40}_{-0.30}$ & 1.46$^{+0.51}_{-0.47}$ & 5.12$^{+1.07}_{-0.87}$ & 4.22$^{+1.10}_{-0.86}$ & 31.48/50 \\ \hline
S11 & 0.839 $\pm$ 0.146 & 4.17 & 0.11$^{+0.26}_{-0.11}$ & 0.92$^{+0.50}_{-0.46}$ & 5.39$^{+1.65}_{-1.21}$ & 5.19$^{+1.78}_{-1.32}$ & 29.14/32 \\ \hline
S12 & 0.646 $\pm$ 0.126 & 4.35 & 0.28$^{+0.32}_{-0.22}$ & 1.99$^{+0.70}_{-0.61}$ & 3.07$^{+1.16}_{-0.70}$ & 2.38$^{+0.87}_{-0.61}$ & 23.93/22 \\ \hline
S13 & 0.852 $\pm$ 0.082 & 13.20 & 0.32$^{+0.14}_{-0.12}$ & 1.74$^{+0.27}_{-0.26}$ & 4.24$^{+0.54}_{-0.48}$ & 3.42$^{+0.50}_{-0.43}$ & 95.74/94 \\ \hline
S14 & 1.091 $\pm$ 0.217 & 2.48 & 0.47 (fixed) & 1.42 $\pm$ 0.34 & 
6.17$^{+1.44}_{-1.20}$ & 5.09$^{+1.43}_{-1.15}$ & 29.18/24 \\ \hline
S15 & 0.818 $\pm$ 0.193 & 2.44 & $<0.25$ & $0.97^{+0.60}_{-0.39}$ & 5.13$^{+2.30}_{-1.57}$ & 5.10$^{+2.30}_{-1.57}$ & 21.08/18 \\ \hline
S16 & 0.917 $\pm$ 0.161 & 3.84 & 0.81$^{+0.48}_{-0.40}$ & 2.65$^{+0.73}_{-0.65}$ & 6.85$^{+5.99}_{-2.35}$ & 3.00$^{+0.83}_{-0.64}$ & 46.46/31 \\ \hline
S17 & 1.048 $\pm$ 0.278 & 1.49 & $<0.20$  & 0.99$^{+0.58}_{-0.43}$ & 5.81$^{+3.27}_{-2.04}$ & 5.77$^{+3.27}_{-2.04}$ & 15.59/14 \\ \hline
S18 & 2.435 $\pm$ 0.405 & 1.52 & 0.68$^{+0.49}_{-0.37}$ & 2.28$^{+0.75}_{-0.66}$ & 14.90$^{+10.23}_{-3.92}$ & 8.51$^{+2.73}_{-1.95}$ & 23.75/33 \\ \hline
S19 & 2.818 $\pm$ 0.417 & 1.66 & 0.85$^{+0.42}_{-0.32}$ & 2.68$^{+0.62}_{-0.55}$ & 20.75$^{+14.82}_{-6.22}$ & 8.74$^{+2.00}_{-1.59}$ & 39.2/40 \\
\bottomrule
\end{tabular}

\raggedright 
\textbf{Notes.} Parameter uncertainties are quoted at the 1$\sigma$ confidence level.  
$^a$ Total count rate in the 0.3--10 keV band.  
$^b$ Intrinsic hydrogen column density.  
$^c$ Photon index.  
$^d$ Luminosity in the 0.5--8 keV band.  
$^e$ C-statistic and degrees of freedom. 
\end{table*}

We have used the X-ray spectral fitting package \texttt{XSPEC}\footnote{\url{https://heasarc.gsfc.nasa.gov/xanadu/xspec/}} version 12.13.0c \citep{Arnaud1999} for spectral analysis. All spectra are fit in the energy range of 0.3--10~keV. For \textit{XMM-Newton}, MOS1, MOS2 and pn spectra are fit simultaneously. In a joint spectral fitting, the multiplicative normalisation constant for the pn detector was fixed to unity, while the constants for the MOS detectors were allowed to vary to account for cross-calibration differences. In observation XM2, ULX-8 lies on the pn chip gap, leading to a flux drop; hence, for fitting purposes, the MOS1 detector's constant value is set to unity for flux estimation.

We model the spectra using absorbed single-component models, namely an absorbed power-law (\texttt{tbabs*tbabs*powerlaw}; \texttt{PL}) and an absorbed multicolour disc blackbody (\texttt{tbabs*tbabs*diskbb}; \texttt{DISKBB}; \citealt{Mitsuda1984}). Spectral fitting is performed in \texttt{XSPEC}, with absorption described by two \texttt{TBABS} components adopting abundances from \citet{Wilms2000} and photoionisation cross sections from \citet{Verner1996}. The first absorption component is fixed to the Galactic hydrogen column density along the line of sight to NGC~4490, $N_{\mathrm{H,Gal}} = 1.81 \times 10^{20}\mathrm{cm^{-2}}$ \citep{Kalberla2005, Doroshenko2024}. The second component represents intrinsic absorption, $N_{\mathrm{H,int}}$, associated with material within the host galaxy and/or the immediate environment of the source. Unless stated otherwise, $N_{\mathrm{H,int}}$ is treated as a free parameter in the fits. The resulting best-fit parameters for the \textit{XMM-Newton} and \textit{Chandra} observations are summarised in Table~\ref{ParameterTable}.

For the \textit{Swift}-\textsc{XRT} observations, only the \texttt{PL} model parameters could be constrained for our dataset (Table~\ref{Swift-Parameter-table}); consequently, \texttt{DISKBB} fits are not reported for these observations. In some fits (C5, C6, C10, C11 with \texttt{DISKBB} and S15, S17 with \texttt{PL}), the best-fit value of $N_{\mathrm{H,int}}$ is consistent with zero. In these cases, we report the upper uncertainty on $N_{\mathrm{H,int}}$ to reflect the allowed range of intrinsic absorption.

A subset of the \textit{Swift}-\textsc{XRT} observations (S7, S8, S9 and S14) lies at the lower end of the available counts and degrees of freedom, limiting the ability of the spectra to independently constrain absorption. Allowing $N_{\mathrm{H,int}}$ to vary yields formally bounded values; however, in all four cases, the propagated uncertainties on the unabsorbed luminosity exceed $100\%$ of the best-fit value. This behaviour indicating that luminosity estimates become dominated by absorption uncertainties in low-count spectra, rendering the resulting luminosity estimates physically uninformative.

Since the primary goal of our analysis is to track luminosity evolution, we therefore adopt a uniform approach and fix $N_{\mathrm{H,int}}$ for these four observations to the mean value derived from the remaining \textit{Swift}-\textsc{XRT} spectra, where absorption is better constrained. This procedure ensures stable estimates of the unabsorbed luminosity while retaining sensitivity to genuine spectral variability through the remaining free parameters.

We find that both the \texttt{PL} and \texttt{DISKBB} models provide statistically acceptable fits to the \textit{XMM-Newton} and \textit{Chandra} spectra. For previously analysed epochs of NGC 4490 ULX-8, the derived parameters are consistent with earlier studies and the adequacy of both models is in agreement with results reported for NGC 4490 ULX-8 as well as for other ULXs such as NGC 253 X-9 and NGC 1313 ULX-2 \citep{Kajava2009}. All best-fit spectra are presented in Appendix~\ref{Spectrums}.

In addition to the \texttt{PL} and \texttt{DISKBB} models, we explored alternative spectral models like \texttt{tbabs*tbabs*diskpbb} (\texttt{DISKPBB}) and \texttt{tbabs*tbabs*cutoffpl} (\texttt{CUTOFFPL}). The \texttt{DISKPBB} model represents a generalized multi-color disk where the local disk temperature scales with radius as \(T(r) \propto r^{-p}\), with \(p\) as a free parameter \citep{Mineshige1994a, Kubota2005}. The standard thin disk (\texttt{DISKBB}) is recovered for \(p = 0.75\), while \(p = 0.5\) corresponds to the slim-disk regime \citep{Abramowicz1988}. Our fits using \texttt{DISKPBB} yielded unconstrained \(p\) values across all the observations except for XM5 (the epoch with the highest post-reduction exposure), where we obtained \(p = 0.71_{-0.10}^{+0.22}\). Fixing \(p = 0.5\) to mimic a slim-disk scenario produced fits that were worse or unchanged in 18 out of 20 (all except XM6 and C8) \textit{Chandra} and \textit{XMM-Newton} observations compared to \(p = 0.75\), favouring a standard thin-disk interpretation.

Similarly, the \texttt{CUTOFFPL} model, which introduces an exponential high-energy cutoff to the power-law, returned unconstrained cutoff energies across all 20 observations. However, spectral turnovers at $\sim$ 5--8 keV have been reported (e.g. \citealt{Roberts2010}, \citealt{Stobbart2006}). We test it by freezing the cutoff at 7.1 keV \citep{Walton2020a}, which allowed the model to fit the spectra adequately, suggesting a possible high-energy cutoff but it remains unmeasurable for NGC 4490 ULX-8 given the current statistical quality of the data.

With these results, we adopt the simpler \texttt{DISKBB} and \texttt{PL} models, which provide stable and well-constrained fits across all observations and are consistent with previous studies of the same ULX. 

The convolution model \texttt{CFLUX} in \texttt{XSPEC} is used to estimate fluxes ($F_{\rm X}$) in the 0.5–8 keV band (to maintain consistency with previous studies). For power-law models, X-ray luminosities are computed assuming isotropic emission as $L_{\rm X} = 4\pi d^{2}F_{\rm X}$, where $d$ is the source distance. For disc models (e.g. \texttt{diskbb}), the luminosity is computed as $L_{\rm X} = 2\pi d^{2} F_{\rm X} / \cos\theta$. In the absence of inclination constraints, we adopt $\theta = 60^\circ$.

Uncertainties are quoted at the 90$\%$ confidence level for one interesting parameter for the \textit{XMM-Newton} and \textit{Chandra} observations (Table~\ref{ParameterTable}) and at the $1\sigma$ level for the \textsc{XRT} observations (Table~\ref{Swift-Parameter-table}).

\subsubsection{F-test}
We tested two-component models such as \texttt{PL+DISKBB}, but they generally did not yield constrained parameters or a statistically significant improvement. To quantify this, we used \texttt{XSPEC}'s \texttt{simftest}\footnote{\url{https://heasarc.gsfc.nasa.gov/docs/software/xspec/manual/node126.html}} \citep{Protassov2002} with 1000 simulations for each comparison (\texttt{PL} vs.\ \texttt{PL+DISKBB} and \texttt{DISKBB} vs.\ \texttt{PL+DISKBB}). The resulting p-values, reported in Table~\ref{ParameterTable}, show that the majority of cases have $p > 0.01$, while some with $p < 0.01$ involve unphysical or unconstrained parameters (see \citealt{Dage2019}), limiting the physical interpretability of those fits. We therefore adopt single-component models (\texttt{PL} or \texttt{DISKBB}) as sufficient to describe the available data for this source.

\section{Results}
\label{results}

\subsection{Long-Term Spectral Variability}

\begin{figure*} 
  \includegraphics[width=7in]{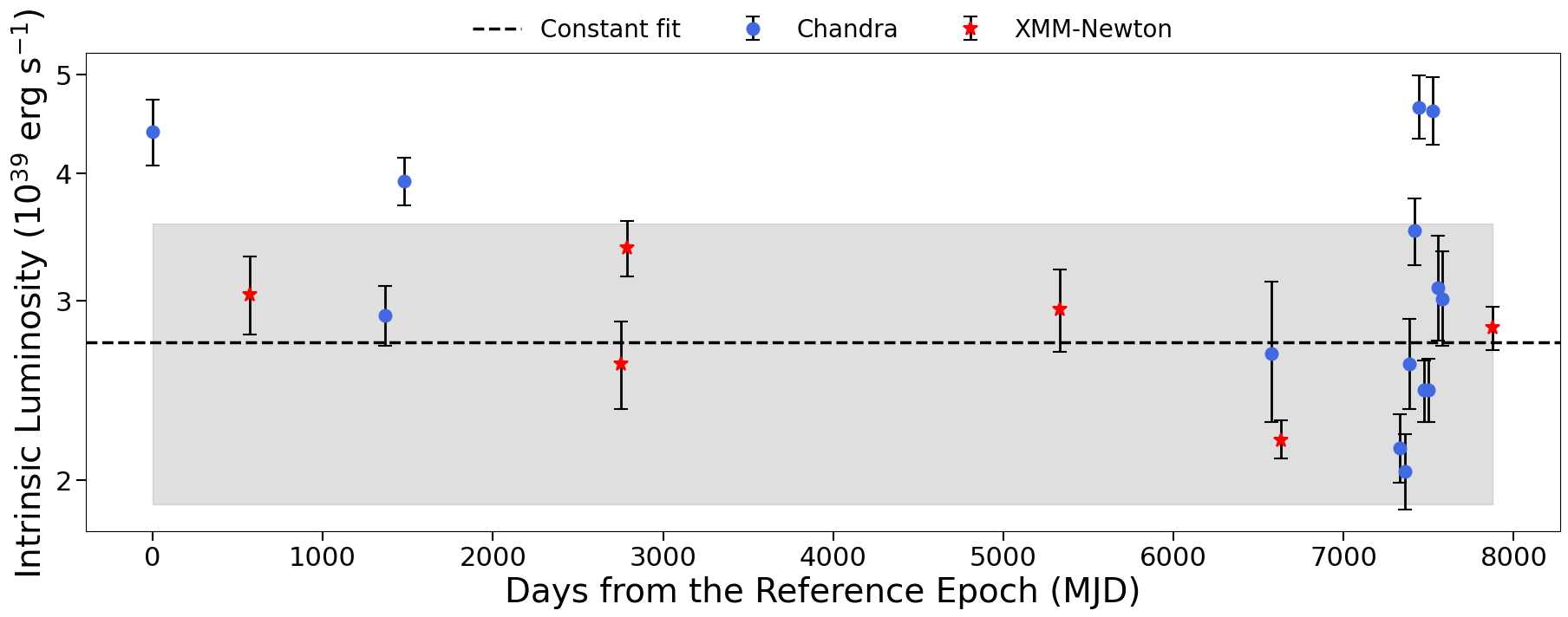}
  \vspace{0.5cm}
  
  \includegraphics[width=7in]{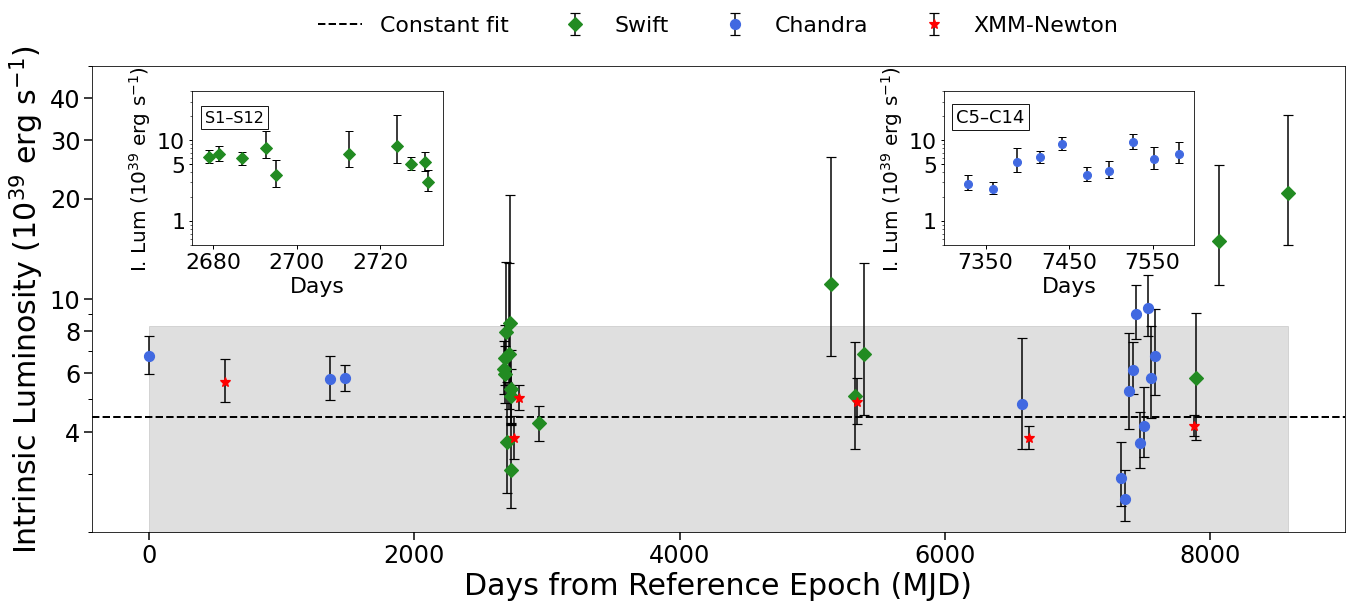}
  \caption{Unabsorbed luminosity vs.\ time graphs for NGC 4490 ULX-8 with the best-fit horizontal line and its RMS error denoted by the shaded region. The oldest observation (C1 with MJD = 51851) is taken as the reference epoch. The Y axis is logarithmically scaled for better visualization. \textit{Upper:} \textit{XMM-Newton} and \textit{Chandra} \texttt{DISKBB} luminosity values. \textit{Lower:} \textit{XMM-Newton}, \textit{Chandra} and \textit{Swift}-\textsc{XRT} \texttt{PL} luminosity values along with the two zoom in plots for the S1--S12 and C5--C14 observation samples.}
  \label{I.Lumin_vs_time}
\end{figure*}

We examine the long-term spectral variability of NGC 4490 ULX-8 across the full observational baseline (Figure~\ref{I.Lumin_vs_time}) using unabsorbed luminosities derived from both spectral \texttt{DISKBB} and \texttt{PL} fits. We fit a constant luminosity model to the data using \texttt{scipy.optimize.curve\_fit}\footnote{\url{https://docs.scipy.org/doc/scipy/reference/generated/scipy.optimize.curve_fit.html}}, shown as a dashed horizontal line in the figure. The root mean square (RMS) error, representing the deviation of the data points from the best-fit line, is indicated by the shaded region. To assess the overall goodness of fit, we compute the reduced chi-squared ($\chi^2$) statistic. 
With \texttt{DISKBB}, the mean unabsorbed luminosity is $L_X = 2.73 \times 10^{39}\ \mathrm{erg\ s^{-1}}$ with an RMS scatter of $0.84\times10^{39}$ erg s$^{-1}$, yielding a poor fit under the assumption of constant luminosity ($\chi^2_\nu=10.36$). For the absorbed power-law model, the mean is $L_X = 4.46 \times 10^{39}\ \mathrm{erg\ s^{-1}}$ with an RMS of $ 3.55 \times 10^{39}\ \mathrm{erg\ s^{-1}}$, still inconsistent with strict constancy ($\chi^2_\nu=1.68$) but markedly better than DISKBB. 

We next examine luminosity changes on intermediate (weeks–months) timescales by focusing on two regularly sampled subsets: the \textit{Chandra} C5–C14 observations and the \textit{Swift}-\textsc{XRT} S1–S12 monitoring. (Figure~\ref{I.Lumin_vs_time}, lower panels). The inferred variability factors are $3.73 \pm 1.04$ for the \textit{Chandra} sample and $2.59 \pm 1.37$ for the \textit{Swift}-\textsc{XRT} sample.

These measurements demonstrate statistically significant variability on both multi-year and week–month timescales, with the inferred amplitude depending on the adopted spectral model.

\subsection{Powerlaw Parameter Correlation Analysis}
\label{Powerlaw Parameter Correlation Study}

\begin{figure*}
  \centering
  \includegraphics[width=0.33\linewidth]{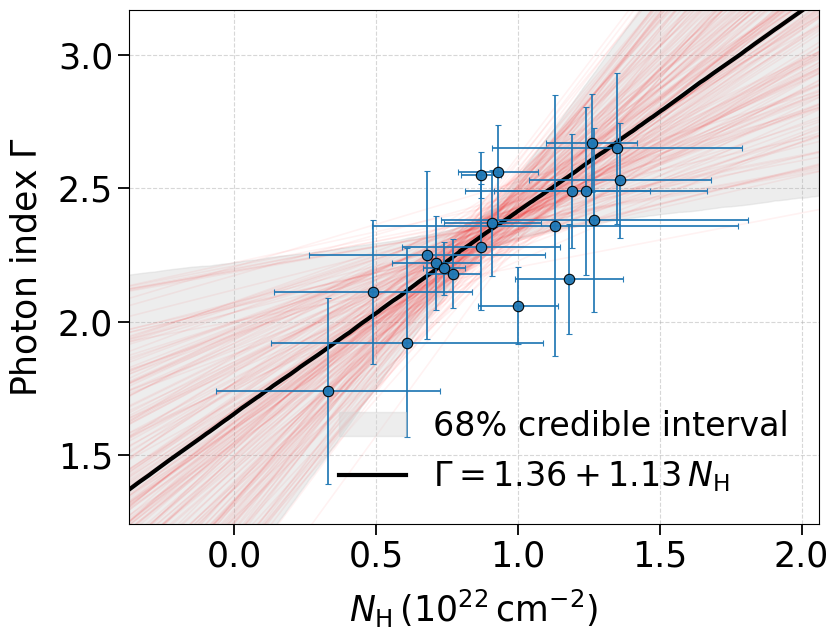}
  \includegraphics[width=0.33\linewidth]{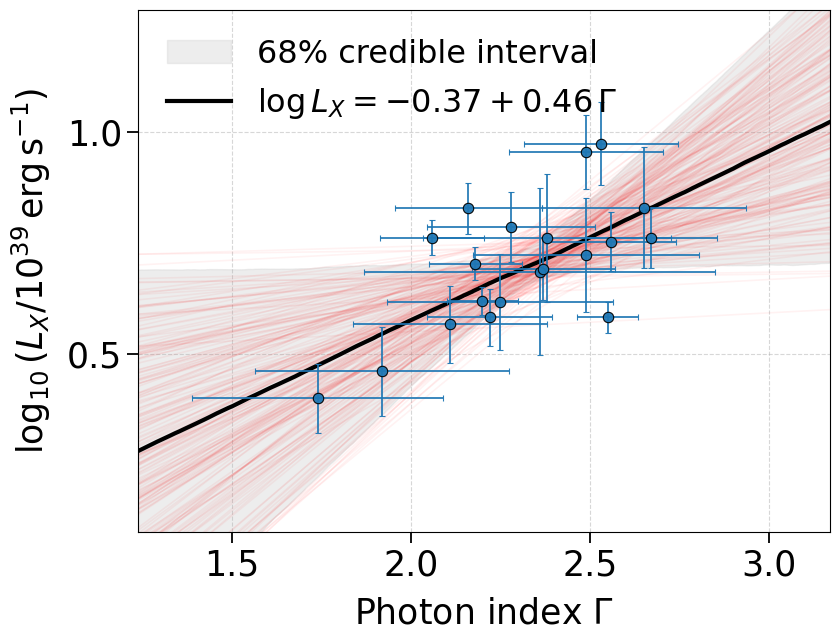}
  \includegraphics[width=0.33\linewidth]{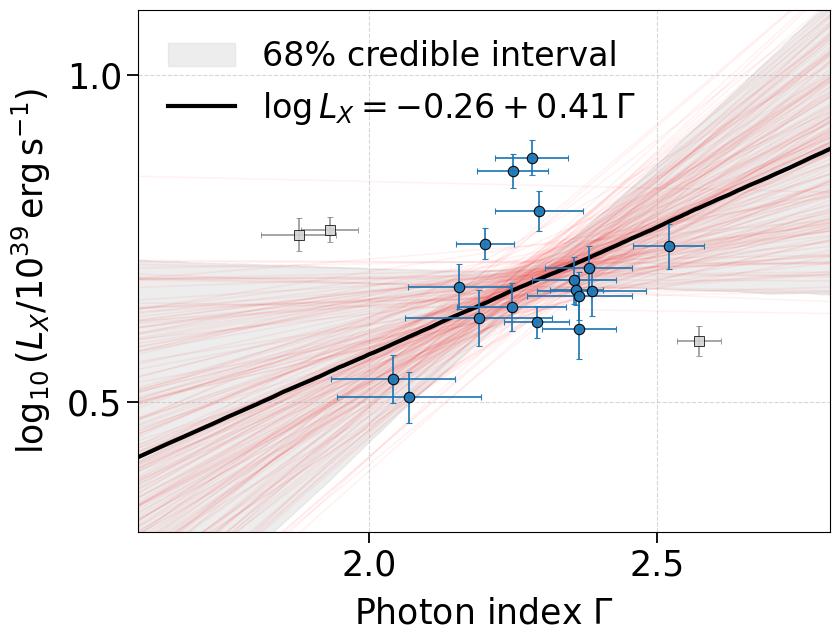}
\caption{%
\texttt{LINMIX} regression results of spectral parameters obtained from
\texttt{PL} fits to \textit{XMM-Newton} and \textit{Chandra} observations.
\textit{Left}: $\Gamma$ versus $N_{\mathrm{H}}$.
\textit{Middle}: $L_{\mathrm{X}}$ versus $\Gamma$ with $N_{\mathrm{H}}$ allowed to vary freely across observations.
\textit{Right}: $L_{\mathrm{X}}$ versus $\Gamma$ with $N_{\mathrm{H}}$ tied across observations; three discrepant observations (XM5, C1, C3; shown in grey) are excluded from the regression to assess the behaviour of the remaining epochs (see Section~\ref{Powerlaw Parameter Correlation Study}). In all panels, the black line shows the posterior median regression, while the grey shaded region indicates the $1\sigma$ credible interval derived from the \texttt{LINMIX} MCMC samples.}
  \label{fig:Linmixplot}
\end{figure*}

\begin{table}
\setlength{\tabcolsep}{8pt}
\centering
\renewcommand{\arraystretch}{1.26}
\caption{Results of Pearson’s linear correlation test and Spearman’s rank correlation test applied for $\Gamma$ vs.\ $N_{\mathrm{H}}$, $L_{\mathrm{X}}$ vs. $\Gamma$ (free $N_{\mathrm{H}}$). Reported are the correlation coefficients ($r$) and the associated two-sided $p$-values.}
\label{correlation test}
\begin{tabular}{@{}cccc@{}}
\toprule
 & & $\Gamma$ vs.\ $N_{\mathrm{H}}$ & $L_{\mathrm{X}}$ vs.\ $\Gamma$ (free $N_{\mathrm{H}}$)  \\
\midrule
Pearson linear test & r-value & 0.76 & 0.58 \\
 & p-value & $9.11 \times 10^{-5}$ & $6.93 \times 10^{-3}$ \\
\midrule 
Spearman rank test & r-value & 0.71 & 0.50 \\
 & p-value & $4.68 \times 10^{-4}$ & $2.41 \times 10^{-2}$ \\
\bottomrule
\end{tabular}
\end{table}

We examine correlations between spectral parameters derived from absorbed power-law (\texttt{PL}) fits by considering three cases: (i) the relation between the photon index $\Gamma$ and intrinsic absorption $N_{\mathrm H}$; (ii) the relation between unabsorbed X-ray luminosity $L_{\mathrm X}$ and $\Gamma$ with $N_{\mathrm H}$ free across epochs; and (iii) the relation between $L_{\mathrm X}$ and $\Gamma$ with $N_{\mathrm H}$ tied across epochs. Throughout this section, the X-ray luminosity is expressed as $\log L_{\mathrm X} \equiv \log_{10}(L_{\mathrm X}/10^{39}\,\mathrm{erg\,s^{-1}})$.

Correlations are first quantified using Pearson’s linear correlation and Spearman’s rank correlation tests, which measure linear and monotonic trends, respectively, without accounting for measurement uncertainties. The resulting correlation coefficients and associated two-sided $p$-values are listed in Table~\ref{correlation test}. A strong positive correlation is found between $\Gamma$ and $N_{\mathrm H}$, with both tests yielding statistically significant results. For the $L_{\mathrm X}$--$\Gamma$ relation with $N_{\mathrm H}$ free, the Pearson and Spearman tests indicate a moderate but statistically significant positive correlation. In the case where $N_{\mathrm H}$ is tied across epochs, Pearson and Spearman tests yield statistically insignificant results, motivating the use of advanced regression methods.

To incorporate measurement uncertainties in both variables, we apply the hierarchical Bayesian regression method \texttt{LINMIX} \citep{Kelly2007}. Since \texttt{LINMIX} does not accommodate asymmetric uncertainties, we approximate asymmetric uncertainties by the mean of the upper and lower error estimates for each parameter. The resulting regression slopes and associated statistical measures are summarised in Table~\ref{linmix table} and shown in Figure~\ref{fig:Linmixplot}.

For all three relations examined, the inferred slopes $\beta$ are positive but not tightly constrained at the $1\sigma$ level (Table~\ref{linmix table}). We therefore characterise the strength of each trend using the posterior probability that the slope is positive, $P(\beta>0)$ and the corresponding equivalent one-sided Gaussian significance, $z_{\mathrm{eq}}$. Values of $P(\beta>0)\simeq0.5$ (or $z_{\mathrm{eq}}\simeq0$) indicate no preference for a positive or negative trend, while progressively larger values reflect increasing evidence for a positive correlation.

In the tied-$N_{\mathrm H}$ analysis, three observations (XM5, C1 and C3; shown in grey in Figure~\ref{fig:Linmixplot}) lie well outside the main locus defined by the majority of the data and exert strong leverage on the inferred regression slope. Notably, these same observations align well with the $\Gamma$--$N_{\mathrm H}$ correlation and also follow the $L_{\mathrm X}$--$\Gamma$ relation in the case where $N_{\mathrm H}$ is allowed to vary freely, indicating that for these epochs the inferred relation is largely driven by absorption-related effects.

Excluding these three observations reveals that the remaining 17 out of 20 epochs define a coherent locus in the $L_{\mathrm X}$--$\Gamma$ plane, with a weak but systematic positive trend. The corresponding \texttt{LINMIX} regression yields a positive slope with $P(\beta>0)=0.82$, although the slope remains only marginally constrained. This demonstrates that the observed $L_{\mathrm X}$--$\Gamma$ trend is characteristic of the majority of the dataset but is not representative of all epochs.

Overall, the results indicate a strong $\Gamma$--$N_{\mathrm H}$ correlation and a weak, population-dominated $L_{\mathrm X}$--$\Gamma$ trend. For most observations, the $L_{\mathrm X}$--$\Gamma$ relation is consistent with reflecting intrinsic spectral behaviour once absorption-related effects are controlled. The physical implications of these correlations are discussed in Section~\ref{Spectral Correlations}.

\begin{table}
\setlength{\tabcolsep}{4pt}
\centering
\renewcommand{\arraystretch}{1.26}
\caption{LINMIX regression results for $\Gamma$ vs.\ $N_{\mathrm{H}}$, $L_{\mathrm{X}}$ vs. $\Gamma$ (free $N_{\mathrm{H}}$) and $L_{\mathrm{X}}$ vs.\ $\Gamma$ (fixed $N_{\mathrm{H}}$). Reported statistic are the best-fit slope $\beta$ and the $1\sigma$ confidence interval (CI) on $\beta$, posterior probability that the slope is positive [$P(\beta > 0)$] and the equivalent Gaussian significance ($z_{\mathrm{eq}}$).}
\label{linmix table}
\begin{tabular}{@{}lccc@{}}
\toprule
Statistic & $\Gamma$ vs.\ $N_{\mathrm{H}}$ & $L_{\mathrm{X}}$ vs.\ $\Gamma$ (free $N_{\mathrm{H}}$) & $ L_{\mathrm{X}}$ vs.\ $\Gamma$ (fixed $N_{\mathrm{H}}$) \\
\midrule
$\beta$ & 1.13 & 0.46 & 0.41 \\
$1\sigma$ CI on $\beta$ & [0.13, 1.49] & [0.01, 0.79] & [-0.03, 0.84] \\
$P(\beta > 0)$ & 0.87 & 0.85 & 0.82 \\
$z_{\mathrm{eq}}$ & 1.15 $\sigma$ & 1.01 $\sigma$& 0.91 $\sigma$ \\

\bottomrule
\end{tabular}
\end{table}
\subsection{Disk Properties and Compact-Object Constraints}

\begin{figure}
    \centering
    \includegraphics[width=2.9in]{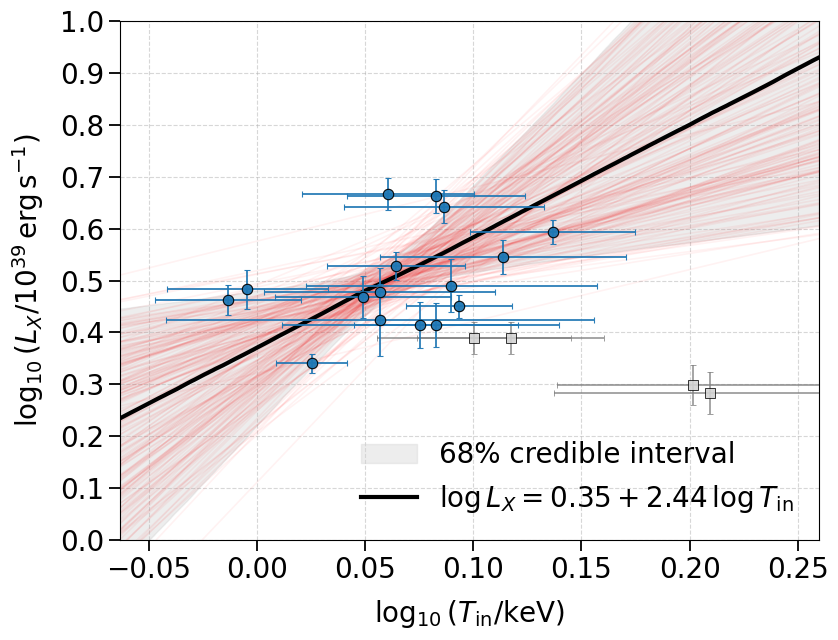}
    \caption{Log–log plot of disk luminosity versus inner disk temperature for ULX-8. The black line shows the posterior median regression, while the grey shaded region indicates the $1\sigma$ credible interval derived from the \texttt{LINMIX} MCMC samples. Grey data points correspond to observations excluded from the fit due to poorly constrained disk parameters.
}
    \label{fig:logL-logT}
\end{figure}

\begin{figure}
    \centering
    \includegraphics[width=2.9in]{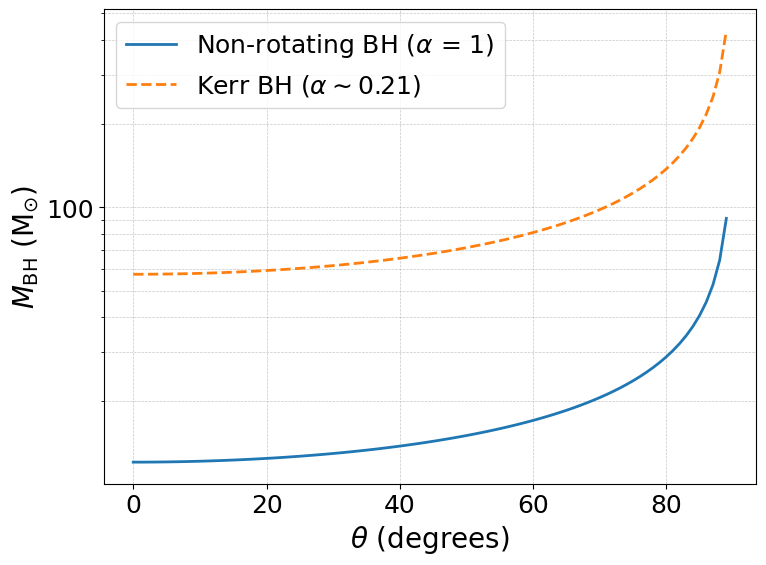} \hspace{0.3cm}
    \caption{\textit{Top}: Variation of black hole mass ($M_{\mathrm{BH}}$) with inclination angle ($\theta$) for non-rotating (Schwarzschild, $a=1$) and rotating (Kerr, $a=0.21$) cases, derived from absorbed multicolour disc blackbody model normalisation.}
    \label{fig:MBH_NS}
\end{figure}

Using parameters from the \texttt{DISKBB} fits, we examine the relation between inner disk temperature ($T_{\mathrm{in}}$) and unabsorbed luminosity ($L_{\mathrm{X}}$), and obtain model-dependent estimates of the characteristic disk radius and compact-object mass.

A \texttt{LINMIX} regression applied to the logarithmic $L_{\mathrm{X}}$–$T_{\mathrm{in}}$ measurements yields a posterior median slope of $\beta = 2.44$, with $P(\beta>0)=0.90$ and $z_{\mathrm{eq}}\simeq$ $1.28\sigma$. The slope is positive but weakly constrained, with a 68$\%$ confidence interval of $[0.51,4.48]$.

In performing the regression analysis, we restrict the sample to observations for which the spectral parameters are reliably constrained. A small subset of observations yields intrinsic absorption values whose best-fit values are driven to the lower bound ($N_{\mathrm{H,int}} = 0$, section~\ref{SA}). In such cases, degeneracies between absorption, continuum shape and disk normalization lead to poorly constrained parameters and increased model dependence. To avoid biasingthe regression by these effects, we exclude these observations from the $L_{\mathrm{X}}$–$T_{\mathrm{in}}$ analysis (Figure~\ref{fig:logL-logT}), where reliable disk parameters cannot be recovered.

Although the \texttt{DISKBB} model is a phenomenological description,  its normalisation parameter (\textit{N}) can be mapped to a characteristic inner disk radius ($R_{\mathrm{in}}$) and an associated compact-object mass ($M_{\mathrm{BH}}$ under standard geometric and spectral assumptions).

We use the highest-quality \textit{XMM-Newton} spectral fits, which have the highest signal-to-noise ratios and degrees of freedom, to derive an average \texttt{DISKBB} normalisation. XMM's larger effective area in the soft X-ray band provides the strongest constraints on the thermal disk component, minimising cross-instrument systematics in the inferred disk parameters.
The relation between $R_{\mathrm{in}}$ and \textit{N} is given by:
\begin{equation}
     R_{\mathrm{in}} = \xi \kappa^2 N^{1/2} (\cos\theta)^{-1/2} d_{10\,\mathrm{kpc}}~\mathrm{km}
\end{equation}
where $\theta$ is the inclination angle of the disk, $d_{10\,\mathrm{kpc}}$ is the source distance from us in 10 kpc units. $\xi$, $\kappa$ are the two correction factors where $\xi$ is the geometric factor and $\kappa$ is the hardening factor for observed color temperature. In the case of a standard accretion disk, $\xi \approx 0.412$ \citep{Kubota1998} and $\kappa \approx 1.7$ \citep{Shimura1995}. Using the $R_{\mathrm{in}}$ values, $M_{\mathrm{BH}}$ can be determined from this equation:
\begin{equation}
     M_{\mathrm{BH}} = R_{\mathrm{in}} c^2/(6G\alpha)
\end{equation}
where $c$ is the speed of light and $G$ is the gravitational constant. $\alpha$ depends on the spin parameter ($a^*$) such that $\alpha \approx 0.21$ for a rapidly rotating (Kerr) BH and $\alpha = 1$ for a non-rotating BH.  
We find $R_{\mathrm{in}} \sim 107 \times (\cos\theta)^{-1/2}$ km and consequently $M_{\mathrm{BH}} \sim 12 \times (\cos\theta)^{-1/2}/\alpha$. Figure~\ref{fig:MBH_NS} shows the dependence of the inferred mass on inclination and spin across the full physically allowed range. 

In the absence of independent constraints on the disk inclination, we adopt a representative value of $\theta = 60^{\circ}$, commonly used in ULX studies (e.g. \citealt{Walton2020a}) and explained in \citet{Urquhart2016}. Under this assumption, we obtain $R_{\mathrm{in}}\sim151$ km and $M_{\mathrm{BH}}\sim81 M_{\sun}$ for a Kerr BH and $\sim17 M_{\sun}$ for a non-rotating BH.

\section{Discussions}

\subsection{Variability, Spectral States and Accretion Regime}
We interpret the observed long- and short-timescale variability of NGC~4490~ULX-8 within the framework of ULX spectral states and accretion regimes, noting that steady emission is disfavoured on multi-year timescales under both spectral models. The inferred variability amplitude depends on the adopted spectral description, with the power-law model yielding higher luminosities and larger scatter than \texttt{DISKBB}. This difference may arise from the model-dependent high-energy extrapolation inherent to the power-law description, as well as from differences in the fitted spectral parameters across epochs.

\citet{Sutton2013} classified ULXs into three spectral regimes: hard ultraluminous, soft ultraluminous and broadened disc, based on spectral morphology, curvature and variability properties. Across all epochs, the X-ray spectra of NGC~4490~ULX-8 are dominated by a smooth, single curved continuum in the 0.3--10~keV band, with a spectral peak at $\sim$1--2~keV and a monotonic decline toward higher energies. No persistent hard excess above $\sim$5--6~keV or distinct soft excess at low energies is observed. Although both \texttt{PL} and \texttt{DISKBB} models provide statistically acceptable fits in individual epochs, two-component models are not required by formal model comparison and remain unconstrained by the available data, indicating intrinsically single-component spectral curvature rather than multiple physically distinct emission components.

These properties, together with disc-dominated fits yielding luminosities of a few $\times 10^{39}\ \mathrm{erg\ s^{-1}}$, place ULX-8 in the broadened-disc regime. The $L_{\mathrm{X}}$–$T_{\mathrm{in}}$ relation provides a consistency check on the accretion flow; however, the inferred slopes are weakly constrained and remain compatible, within uncertainties, with both thin-disc ($L_{\mathrm{X}}\propto T_{\mathrm{in}}^{4}$; \citealt{Shakura1973}) and slim-disc ($L_{\mathrm{X}}\propto T_{\mathrm{in}}^{2}$; \citealt{Watarai2001}) scalings, and therefore do not uniquely distinguish between accretion regimes. Considered together with the persistent spectral morphology, luminosity range, and absence of state transitions, the results are consistent with the source remaining in a single accretion regime, with accretion near the Eddington limit.

The observed spectral variability is therefore most naturally attributed to changes in the curvature of the dominant continuum rather than transitions between distinct accretion states. Several mechanisms could contribute to such variability on multi-year timescales, including modulation of the mass transfer rate from the donor star, potentially driven by stellar activity cycles \citep{Applegate1992, Kotze2012}, or changes in the accretion geometry at near- or super-Eddington rates, such as variations in wind strength, funnel opening angle, or line-of-sight obscuration \citep{Middleton2015, Gurpide2021}.

On shorter (week–month) timescales, luminosity variations of up to a factor of $\sim$4 remain consistent with persistent to moderately variable behaviour reported in other ULXs \citep{Earnshaw2025}, such as M51~ULX-8 and NGC~4490~ULX-4 and in high-mass X-ray binaries.

\subsection{Spectral Correlations and Implications for the Compact Object}
\label{Spectral Correlations}

Correlations between spectral parameters derived from absorbed power-law fits have been widely reported in ULXs and used as diagnostics of spectral variability, though these are potentially affected by model degeneracies \citep[e.g.][]{Kajava2009}. 

When intrinsic absorption is allowed to vary freely across epochs in NGC~4490~ULX-8, a positive $L_{\mathrm{X}}$--$\Gamma$ correlation is observed, accompanied by a strong $\Gamma$--$N_{\mathrm{H}}$ correlation. This covariance is expected in absorbed power-law models: increases in $N_{\mathrm{H}}$ suppress soft photons and are naturally compensated by steeper photon indices. Because unabsorbed luminosity depends sensitively on both parameters, this coupling can artificially enhance the apparent $L_{\mathrm{X}}$--$\Gamma$ correlation when absorption is removed using the fitted values.

When intrinsic absorption is instead tied across epochs, the behaviour changes markedly. A small subset of observations deviates from the main $L_{\mathrm{X}}$--$\Gamma$ locus, but the remaining majority of epochs define a weak, statistically significant positive trend. This indicates that while absorption-related effects dominate a few epochs, most observations retain a consistent spectral response when absorption-driven variability is controlled. The modest significance implies any intrinsic correlation is weak and should be interpreted cautiously.

If this weak trend reflects genuine spectral evolution, it can be interpreted within near- or super-Eddington accretion frameworks. Increasing mass accretion rates drive radiatively launched winds that modify the effective photosphere of the flow. Enhanced reprocessing in the wind or funnel increases spectral curvature, producing progressively softer observed spectra as luminosity rises. In this picture, variations in fitted $\Gamma$ primarily trace changes in continuum curvature rather than a simple power-law slope.

The observed positive $L_{\mathrm{X}}$--$\Gamma$ trend of softer spectra at higher luminosities, has been reported for several ULXs including NGC~1313~ULX-1, Holmberg~IX~ULX-1, Holmberg~II~ULX-1, and NGC~5204~ULX-1, while others such as NGC~253~X-2 and NGC~1313~ULX-2 exhibit anti-correlations. This diversity demonstrates that luminosity--spectral-index relations are not universal across the ULX population.

NGC~4490~ULX-8 lies closer to the parameter space typically occupied by BH ULXs than to that associated with NS ULXs. With hardness ratios $\mathrm{HR}\approx0.8-3$ and photon indices $\Gamma\approx0.9-2.7$, it falls outside the harder spectral region preferentially occupied by NS ULXs, where magnetically channelled accretion produces characteristically hard continua \citep{Gurpide2021}. Instead, the spectral properties of ULX-8 are consistent with disk-dominated emission expected from super-critical disc and wind geometries \citep{Gladstone2009,Sutton2013,Kaaret2017}, favouring an interpretation in which the compact object is a stellar-mass BH.

\section{Conclusions}

We have presented a comprehensive X-ray spectral and variability study of the ultraluminous X-ray source NGC~4490~ULX-8 using 14 \textit{Chandra}, 6 \textit{XMM-Newton} and 19 selected \textit{Swift}-\textsc{XRT} observations obtained between 2000 and 2024. Our main conclusions are as follows:

\begin{enumerate}

\item \textbf{Spectral properties and luminosity range:}
Across all epochs, the source spectra are adequately described by absorbed power-law and multicolour disc blackbody models. The inferred unabsorbed luminosities span $\sim(3$--$20)\times10^{39}$~erg~s$^{-1}$ for power-law fits and $\sim(2$--$5)\times10^{39}$~erg~s$^{-1}$ for disc-dominated fits, firmly placing ULX-8 in the ultraluminous regime.

\item \textbf{Variability behaviour:}
ULX-8 exhibits significant variability over the $\sim$24-year observational baseline, while remaining comparatively stable within individual observations. On week--month timescales, variability factors of $3.73\pm1.04$ and $2.59\pm1.37$ are measured from regularly sampled \textit{Chandra} and \textit{Swift}-\textsc{XRT} data, respectively, placing the source among persistent to moderately variable ULXs. No robust short-term periodicities are detected, although mild fractional variability is present in a small subset of observations.

\item \textbf{Spectral state and accretion regime:}
The X-ray spectra across all epochs are characterised by a smooth, single curved continuum peaking at $\sim$1--2~keV, with no persistent hard excess or distinct soft component. Combined with the observed luminosity range and moderate variability, these properties are consistent with the broadened-disc ultraluminous regime. No evidence is found for transitions between canonical hard and soft spectral states.

\item \textbf{Spectral correlations:}
A positive $L_{\rm X}$--$\Gamma$ correlation is observed across all epochs when $N_{\rm H}$ is allowed to vary freely, accompanied by a strong $\Gamma$--$N_{\rm H}$ correlation. When $N_{\rm H}$ is tied across epochs, a weak but statistically significant positive $L_{\rm X}$--$\Gamma$ trend persists for the majority of epochs, indicating that absorption-related effects influence but do not fully account for the observed spectral variability.

\item \textbf{Disc temperature and compact-object constraints:}
The $L_X$–$T_{\rm in}$ relation yields a positive but weakly constrained slope due to substantial temperature uncertainties, remaining compatible with both thin-disc and slim-disc scalings. Disc-based mass estimates derived from the highest-quality \textit{XMM-Newton} spectra are inherently model-dependent, but yield illustrative values spanning $\sim$17--81~M$_\odot$, depending on the assumed black hole spin.

\item \textbf{Nature of the compact object:}
The spectral hardness and photon index of ULX-8 place it outside the parameter space typically occupied by NS ULXs. Combined with the mass estimates and the observed spectral shape, the source is consistent with a stellar-mass BH accretor operating in the broadened-disc ultraluminous regime operating at or near the Eddington limit.

\end{enumerate}

\section*{Acknowledgements}

We thank the anonymous referee for their constructive comments, which helped improve the manuscript. We thank the Raman Research Institute (RRI) for support during the initial period of this work under the Visiting Student Program (VSP), which provided funding and the opportunity to carry out research. TV acknowledges Ganiv Kaur for numerous useful discussions. AB acknowledges the financial support from SERB (SB/SRS/2022-23/124/PS) and is grateful to the Royal Society, United Kingdom.
This work is based on observations obtained with \textit{XMM-Newton}, an ESA science mission with instruments and contributions directly funded by ESA Member States and NASA. This research has made use of data obtained from the \textit{Chandra} Data Archive and software provided by the \textit{Chandra} X-ray Center (CXC) in the application package \texttt{CIAO}. We also acknowledge the use of public data from the \textit{Swift} data archive.

\section*{Data Availability}

 All the data sets described in Section~\ref{Obs} of this paper are publicly available. The \textit{XMM-Newton} and \textit{Swift}-XRT data can be downloaded from NASA’s High Energy Astrophysics Science Archive Research Center (HEASARC; \href{https://heasarc.gsfc.nasa.gov/docs/archive.html}{https://heasarc.gsfc.nasa.gov/docs/archive.html}). \textit{XMM-Newton}'s data can also be accessed from ESA’s \textit{XMM-Newton} Science Archive (XSA; \href{https://www.cosmos.esa.int/web/xmm-newton/xsa}{https://www.cosmos.esa.int/web/xmm-newton/xsa}). The \textit{Chandra} data are available from the \textit{Chandra} Data Archive (\href{https://asc.harvard.edu/cda/}{https://asc.harvard.edu/cda/}).



\bibliographystyle{mnras}
\bibliography{Main-Paper} 




\appendix

\section{Best-fit Spectra}
\label{Spectrums}

\begin{figure*}
  \centering
  \includegraphics[width=2.1in]{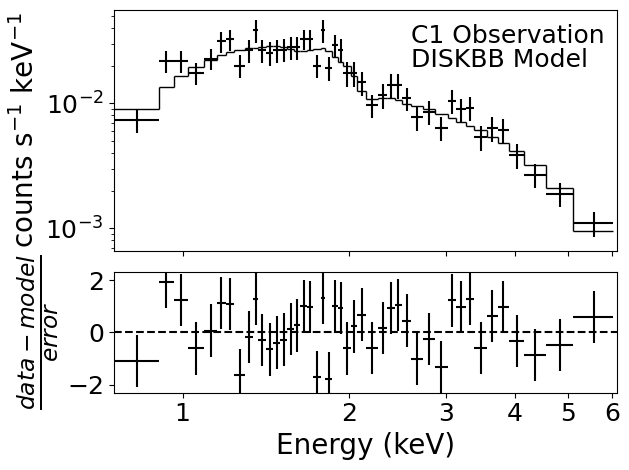} \hspace{0.3cm}
  \includegraphics[width=2.1in]{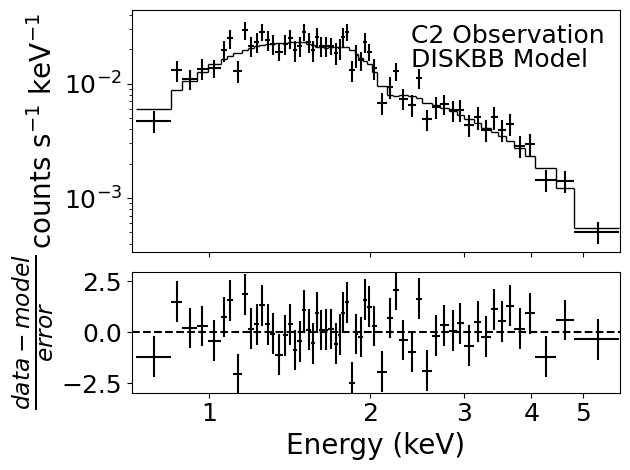} \hspace{0.3cm}
  \includegraphics[width=2.1in]{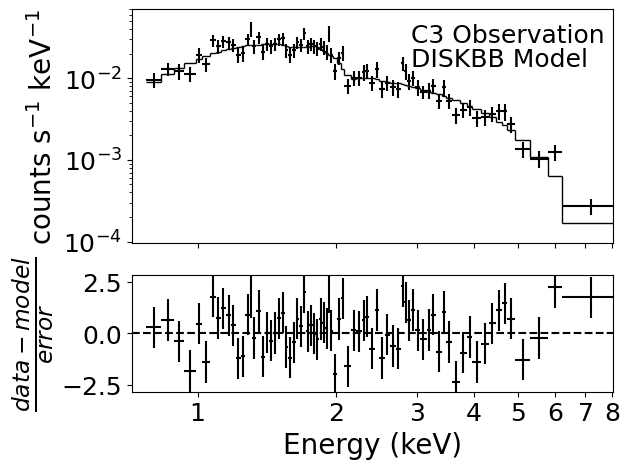} \vspace{0.3cm} \\
  \includegraphics[width=2.1in]{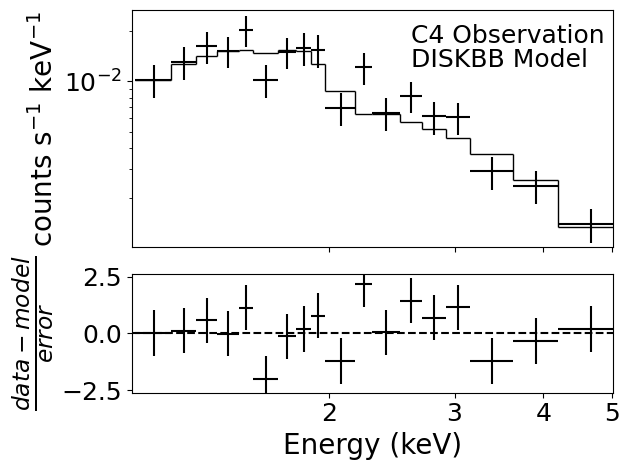} \hspace{0.3cm}
  \includegraphics[width=2.1in]{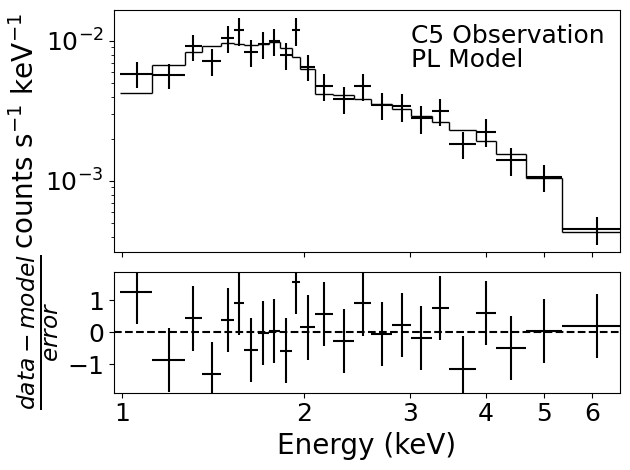} \hspace{0.3cm}
  \includegraphics[width=2.1in]{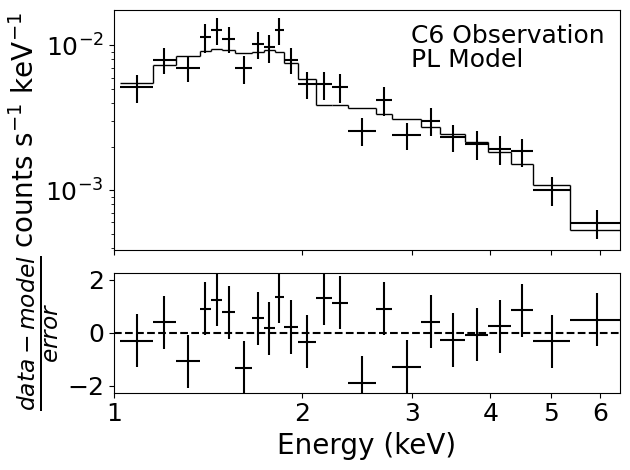} \vspace{0.3cm} \\
  \includegraphics[width=2.1in]{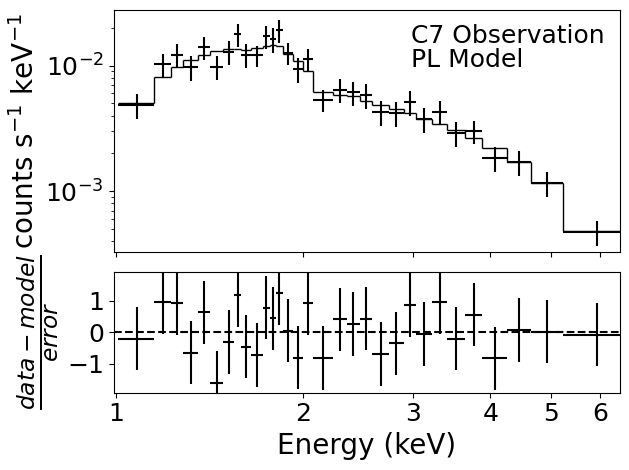} \hspace{0.3cm}
  \includegraphics[width=2.1in]{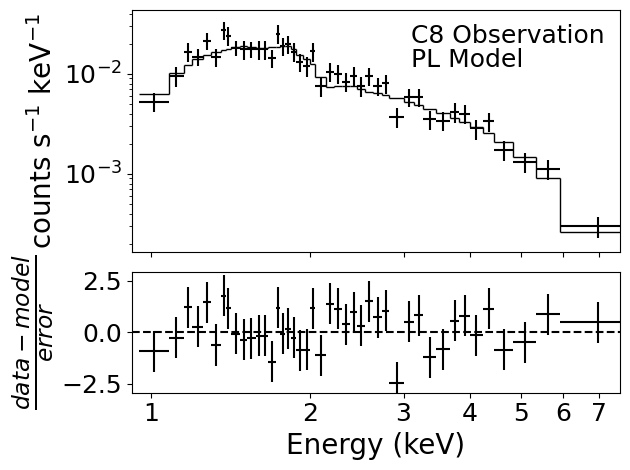} \hspace{0.3cm}
  \includegraphics[width=2.1in]{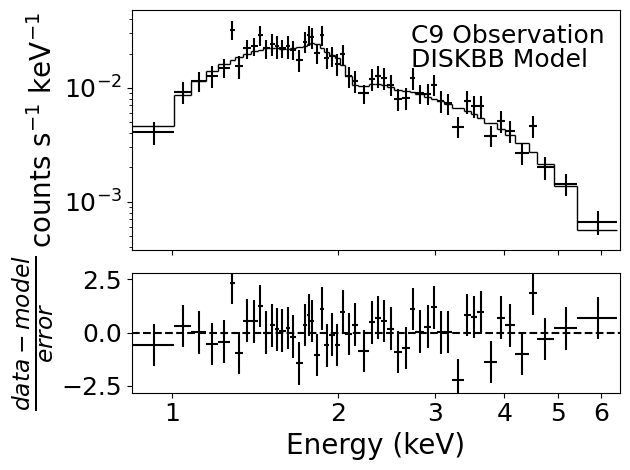} \vspace{0.3cm} \\
  \includegraphics[width=2.1in]{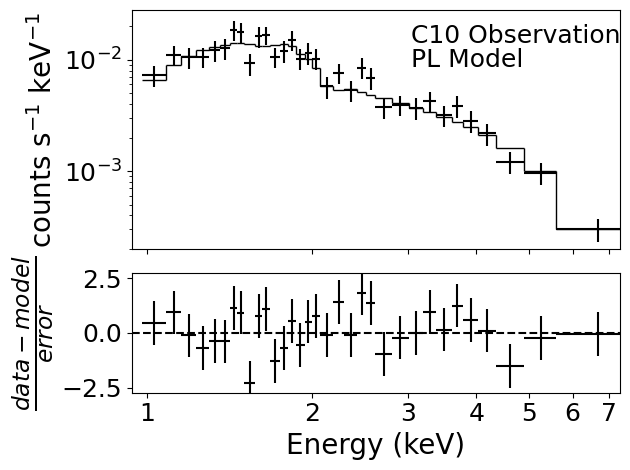} \hspace{0.3cm}
  \includegraphics[width=2.1in]{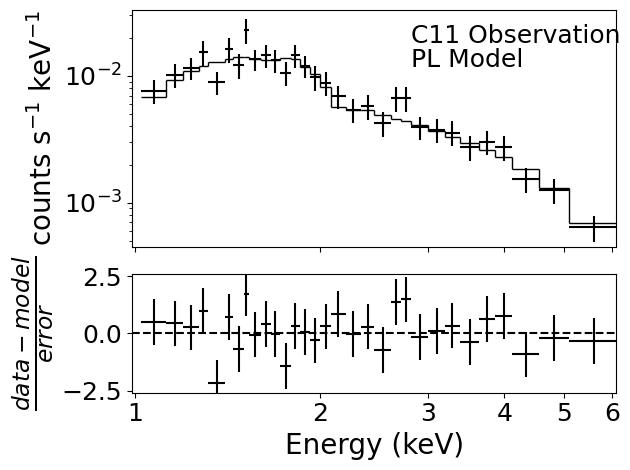} \hspace{0.3cm}
  \includegraphics[width=2.1in]{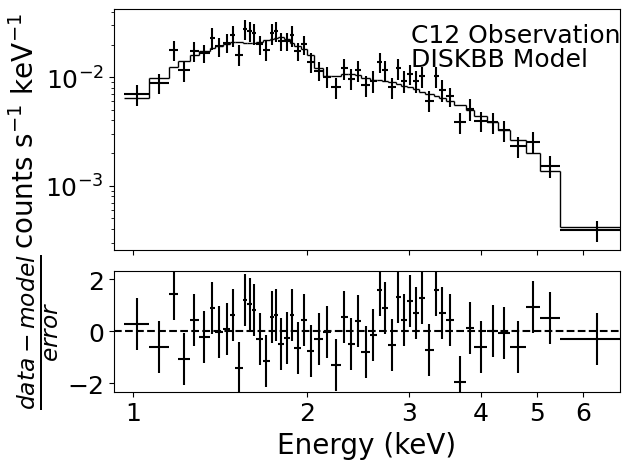} \vspace{0.3cm} \\
  \includegraphics[width=2.1in]{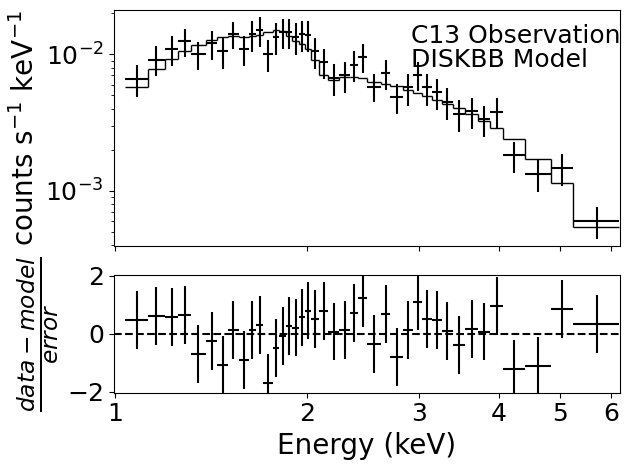} \hspace{0.3cm}
  \includegraphics[width=2.1in]{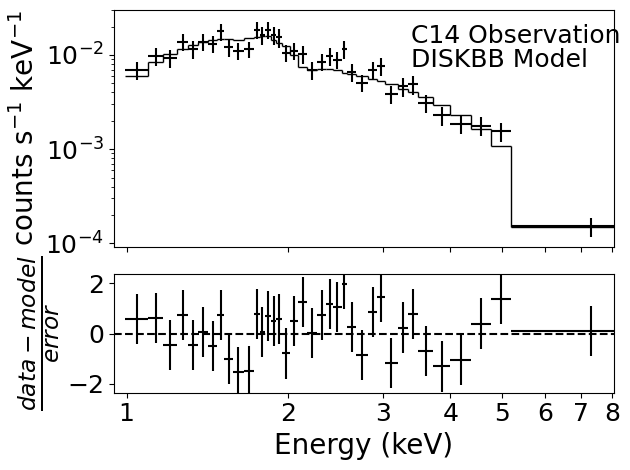} \vspace{0.3cm}

  \caption{Best-fitting spectra of all \textit{Chandra} observations of NGC~4490~ULX-8 obtained with the ACIS instrument. Each panel shows the observed spectrum (data points with error bars) fitted with best fit of either a multicolour disc blackbody (\texttt{DISKBB}) or a power-law (\texttt{PL}) model, along with the corresponding residuals in the lower sub-panel.}
  \label{fig:chandra_spectra}
\end{figure*}

\begin{figure*}
  \centering
  \includegraphics[width=2.2in]{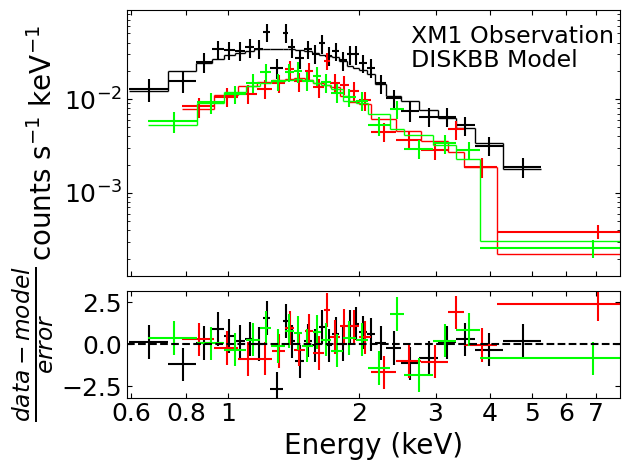} \hspace{0.3cm}
  \includegraphics[width=2.2in]{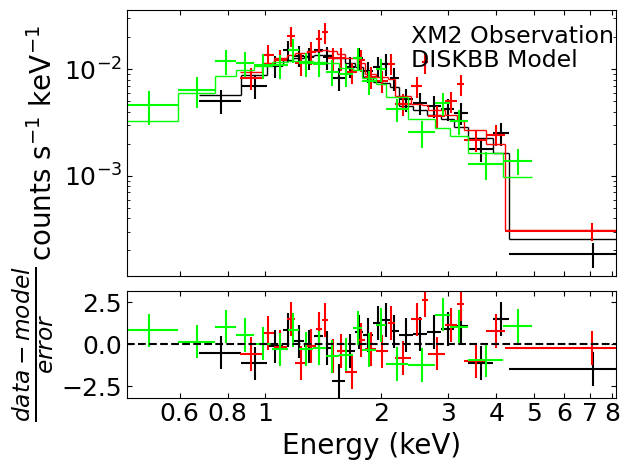} \hspace{0.3cm}
  \includegraphics[width=2.2in]{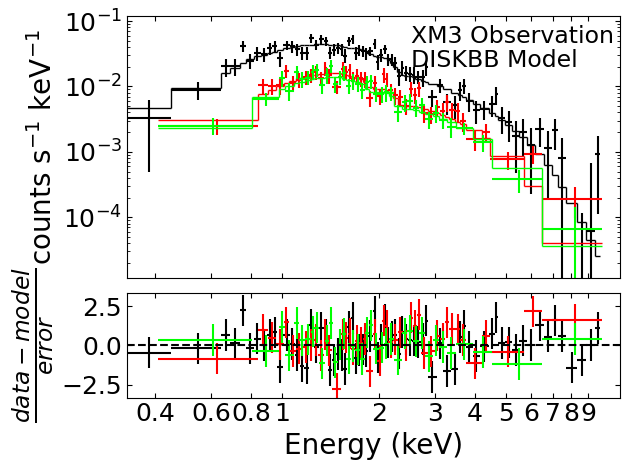} \vspace{0.3cm} \\
  \includegraphics[width=2.2in]{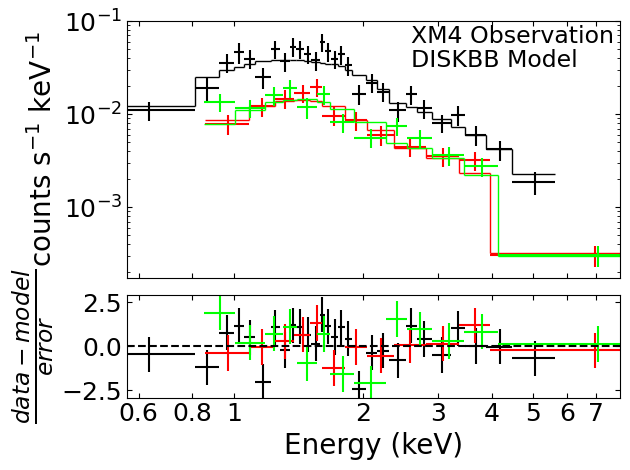} \hspace{0.3cm}
  \includegraphics[width=2.2in]{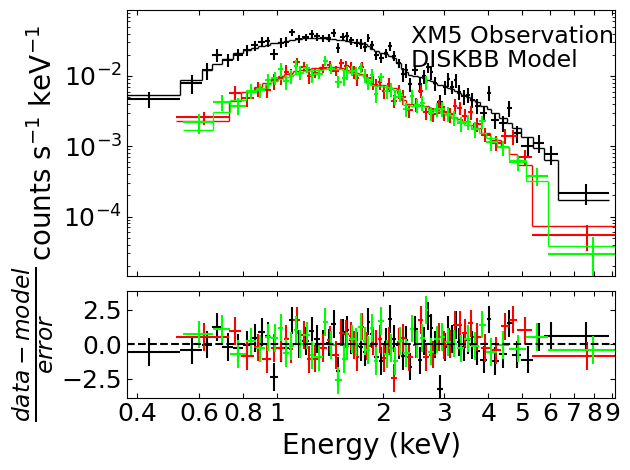} \hspace{0.3cm}
  \includegraphics[width=2.2in]{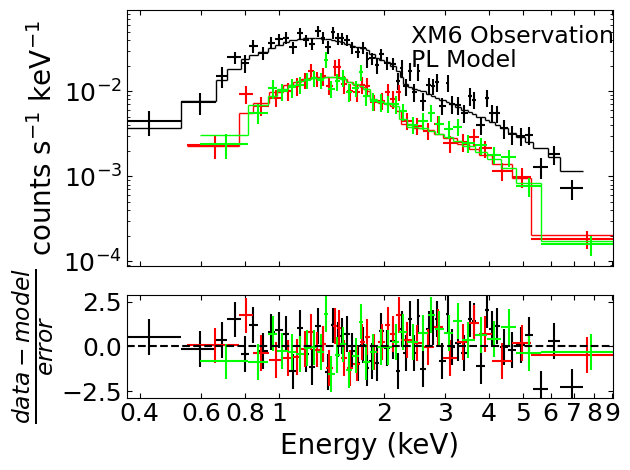} \vspace{0.3cm}

  \caption{Best-fitting spectra of NGC~4490~ULX-8 from \textit{XMM-Newton} observations. Data from the EPIC-pn, MOS1 and MOS2 detectors are shown in black, red and green, respectively. Each panel shows the observed spectrum (data points with error bars) fitted with best fit of either a multicolour disc blackbody (\texttt{DISKBB}) or a power-law (\texttt{PL}) model, along with the corresponding residuals in the lower sub-panel.}
  \label{fig:xmm_spectra}
\end{figure*}

The best-fit spectra for the 14 \textit{Chandra} observations are shown in Figure~\ref{fig:chandra_spectra} and those for the 7 \textit{XMM-Newton} observations are shown in Figure~\ref{fig:xmm_spectra}.

\bsp	
\label{lastpage}
\end{document}